\documentclass[12pt,a4paper,titlepage,oneside]{report}

\usepackage{times}
\usepackage[sort&compress]{natbib}
\usepackage[utf8]{inputenc}
\usepackage{t1enc}
\usepackage[english]{babel}
\usepackage{amsmath}
\usepackage{amssymb}
\usepackage{ae}
\usepackage[font=small,labelfont=bf,up,hang]{caption}
\usepackage{color}
\usepackage{array}
\usepackage{longtable}
\usepackage{calc}
\usepackage{multirow}
\usepackage{hhline}
\usepackage{ifthen}
\usepackage{amssymb}
\usepackage{bm}
\usepackage{ifpdf}
\usepackage{icomma}
\usepackage{verbatim}
\usepackage{setspace}
\usepackage[pdftex]{graphicx}
\usepackage[pdftex]{hyperref} 
\hypersetup{colorlinks=true,
	pdfstartview=FitV,
	linkcolor=blue,
	citecolor=red,
	urlcolor=blue,
	pdfauthor={HORVATH Laszlo, hovath.laszlo@reak.bme.hu},
	pdfsubject={MSc 2015},
	pdftitle={Analysis of fast ion induced instabilities in tokamak plasmas}}
\DeclareGraphicsExtensions{.pdf}

\newcommand{\dt}{\mathrm{d}t}
\newcommand*\rfrac[2]{{}^{#1}\!/_{#2}}

\begin{document}

\begin{titlepage}
\begin{center}

\vspace{-8 cm}
\centerline{\resizebox{80mm}{!}{\includegraphics{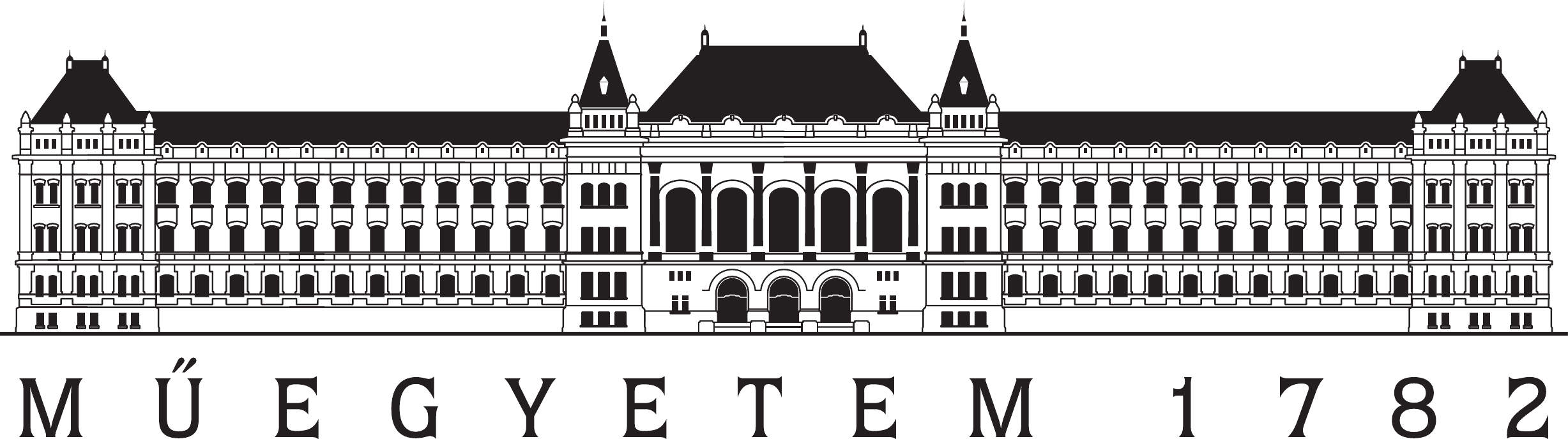}}}
\vspace{1.5 cm}
\textsc{\LARGE{MSc thesis}}\\
\vspace{2.5 cm}
\begin{spacing}{1.3}
\textsc{\LARGE{Analysis of fast ion induced instabilities in tokamak plasmas}}\\
\end{spacing}
\vspace{1.0 cm}
\textbf{\Large{L\'{a}szl\'{o} Horv\'{a}th}} \\
\vspace{3.0 cm}
\hspace{-2.3 cm}\begin{tabular}{p {3 cm} r p{0.1 cm} l}
& {\normalsize Supervisors:} & & {\normalsize \bf Dr. Gergely Papp}\\
& & & {\normalsize \it postdoctoral researcher}\\
& & & {\normalsize Max-Planck/Princeton Center for Plasma Physics}\\
& & & \\
& & & {\normalsize \bf Dr. Gerg\H{o} Pokol}\\
& & & {\normalsize \it associate professor}\\
& & & {\normalsize BME Institute of Nuclear Techniques}\\
& & & {\normalsize Budapest, Hungary}\\
\end{tabular}\\
\vspace{1.9 cm}
{\large \bf Budapest University of Technology and Economics}\\
\vspace{0.2 cm}
{\large \bf 2015}
\end{center}
\end{titlepage}

{\small
\noindent \textsc{Analysis of fast ion induced instabilities in tokamak plasmas}\\
László Horváth\\
Institute of Nuclear Techniques\\
Budapest University of Technology and Economics}
\vspace{0.7 cm}

\noindent {\bf \large Abstract}
\vspace{0.3 cm}

\noindent
Nuclear fusion is a promising energy source of the future.
One possible way to achieve reliable energy production from fusion reactions is to confine the high temperature deuterium-tritium fuel by magnetic fields.
In magnetic confinement fusion devices like tokamaks, it is crucial to confine the high energy fusion-born helium nuclei ($\alpha$-particles) to maintain the energy equilibrium of the plasma.
However, super-thermal energetic ions can excite various instabilities which can lead to their enhanced radial transport.
Consequently, these instabilities may degrade the heating efficiency and they can also cause harmful power loads on the plasma-facing components of the device.
Therefore, the understanding of these modes is a key issue regarding future burning plasma experiments.

One of the main open questions concerning energetic particle (EP) driven instabilities is the non-linear evolution of the mode structure.
In this thesis, I present my results on the investigation of $\beta$-induced Alfv\'{e}n eigenmodes (BAEs) and EP-driven geodesic acoustic modes (EGAMs) observed in the ramp-up phase of off-axis neutral beam injection heated plasmas in the ASDEX Upgrade tokamak.
These modes were well visible on several line-of-sights of the soft X-ray cameras which made it possible to analyse the time evolution of their spatial structure.

In order to investigate the radial structure, the mode amplitude has to be determined on different line-of-sights.
I developed an advanced amplitude reconstruction method which can handle the rapidly changing mode frequency and the low signal-to-noise ratio.
This method is based on short time Fourier transform which is widely applied in the thesis, because it is ideal to investigate the time evolution of transient wave-like phenomena.

The radial structure analysis showed that in case of the observed downward chirping BAEs the changes in the radial eigenfunction were smaller than the uncertainty of the measurement, while in case of rapidly upward chirping EGAMs the analysis shows shrinkage of the mode structure.
These experimental results are shown to be consistent with the corresponding theory.

\thispagestyle{empty}
\clearpage

{\small 
\noindent \textsc{Gyors ionok hajtotta hullámok vizsgálata tokamak plazmákban}\\
Horváth László\\
Nukleáris Technikai Intézet\\
Budapesti Műszaki és Gazdaságtudományi Egyetem}
\vspace{0.7 cm}

\noindent {\bf \large Kivonat}
\vspace{0.3 cm}

\noindent
Jövőnk ígéretes energiaforrása a nukleáris fúzió, mely megvalósításának egyik lehetséges módja mágneses összetartású berendezés alkalmazása.
Ilyen berendezés a tokamak, melyben a deutérium-trícium fúziós reakció során keletkező nagy energiájú hélium atommagok ($\alpha$-részecskék) összetartása különösen fontos a plazma energiaegyensúlyának fenntartása szempontjából.
Ezek a termikusnál jóval nagyobb energiával rendelkező ionok azonban olyan plazmainstabilitásokat kelthetnek, melyek felgyorsíthatják a gyors ionok radiális transzportját.
Ennek következményeként csökken az $\alpha$-részecskék fűtési hatékonysága, továbbá a kiszóródó gyors ionok súlyosan károsíthatják a plazmát körülvevő falelemeket.
Emiatt a gyors ionok által hajtott instabilitások megértése létfontosságú a jövő energiatermelő fúziós reaktorai szempontjából.

A kutatási terület egyik máig nyitott kérdése a gyors ionok által hajtott módusok szerkezetének nem-lineáris időfejlődése.
Jelen dolgozatban az ASDEX Upgrade tokamak semleges atomnyaláb által fűtött plazmáinak kezdeti szakaszában megfigyelt $\beta$-hajtott Alfvén sajátmódusokkal ($\beta$-induced Alfv\'{e}n eigenmode, BAE) és gyors részecskék által hajtott geodézikus akusztikus módusokkal (energetic particle driven geodesic acoustic mode, EGAM) kapcsolatos kísérleti eredményeimet mutatom be.
Ezen módusok jól megfigyelhetőek a lágy-röntgen diagnosztika számos csatornáján, mely segítségével a módusok radiális struktúrájának időfejlődése vizsgálható.

A radiális struktúra tanulmányozásához a módus amplitúdó meghatározására van szükség a lágy-röntgen diagnosztika különböző látóvonalain.
Ennek érdekében kifejlesztettem egy amplitúdó rekonstrukciós eljárást, amely segítségével kezelhető a módus gyorsan változó frekvenciája és az alacsony jel-zaj viszony.
Ez a módszer rövid idejű Fourier-transzformáción alapul, melyet széles körben használtam a munkám során, mivel kiváló tranziens jelek elemzésére.

Vizsgálataim megmutatták, hogy a megfigyelt BAE-k esetében a módus radiális struktúrájának változása kisebb a mérés bizonytalanságánál.
Ezzel szemben az EGAM-ok esetében a módus struktúra összehúzódása figyelhető meg.
Ezek a kísérleti eredmények összhangban vannak a jelenséget leíró elmélettel.
\thispagestyle{empty}
\clearpage

\noindent {\bf \large A szakdolgozat kiírása}

\noindent{Azonosító: DM-2014-44}

\vspace{0.3 cm}
\noindent Önfenntartóan működő szabályozott magfúziós reaktor működése szempontjából kulcsfontosságú a termikusnál jóval nagyobb energiájú ionok összetartása. Ezen ionok rezonáns részecske-hullám kölcsönhatáson keresztül különböző hullámokat tudnak hajtani, mely hullámok a részecskék megnövekedett veszteségét okozzák, ezáltal rontva a plazmafűzés hatékonyságát. A hallgató elsődleges feladata a németországi ASDEX Upgrade tokamakon megfigyelt, gyors részecskékhez köthető különböző Alfvén-sajátmódusok, ezen belül a gyors frekvenciaváltozással jellemezhető ún. ,,csörpölő'' (az angol ,,chirp'' szóból) módusok vizsgálata. Nyitott kérdés, hogy a csörpölő módusok radiális sajátfüggvénye hogyan változik a csörp ideje alatt: a mérési bizonytalanságon belül állandó marad, vagy szignifikánsan változik? Ez az információ kulcsfontosságú a módusok megértése, és későbbi prediktív modellezése szempontjából. A hallgatónak ki kell dolgoznia egy eljárást amellyel a csörpölő módusok sajátfüggvényének időbeli változása vizsgálható, és a változás / nem változás ténye a mérési bizonytalanságon belül szignifikánsan kimutatható. A vizsgálat során elsődlegesen felhasználandó adatok az ASDEX Upgrade tokamakon rendelkezésre álló mágneses diagnosztikák jelei, a vonalintegrált lágy röntgen sugárzás mérése illetve az elektron-ciklotron kibocsájtáson alapuló lokalizált hőmérsékletmérés. Szükség esetén más diagnosztikák is bevonhatók a vizsgálatba. A feladat során meg kell ismerni a diagnosztikák működését, ki kell dolgozni egy eljárást a radiális sajátfüggvény időbeli változásának mérésére és a mérési bizonytalanságok becslésére. A mérési eredmények függvényében, azok mélyebb megértéséhez szükségessé válhat a vizsgált módusok szimulációja és szintetikus diagnosztikák alkalmazása. Ezen szimulációs programok megírása a hallgatónak nem feladata, a célra jól működő programcsomagok állnak rendelkezésre a témavezető intézetében. Az elemzések végrehajtásához segítséget nyújt a BME NTI-ben fejlesztett NTI Wavelet Tools programcsomag. A programcsomag több különböző elemző algoritmus mellett jól működő kísérleti adatbeolvasó-, manipuláló- és ábrázoló képességekkel rendelkezik, ezáltal célszerű a vizsgálathoz a már meglévő programokat felhasználni, szükség esetén azokat módosítani. Az NTI Wavelet Tools nagy része IDL programnyelven íródott, ezért ennek a nyelvnek ismerete vagy szükséges előfeltétel, vagy megtanulandó, mivel a hallgató feladata lesz a szükséges változtatásokat is elvégezni. A szakdolgozat témájához kötődő kutatómunka nemzetközi együttműködésben zajlik, eseti külföldi kiutazásokkal. Magas szintű angol nyelvtudás, alapos programozói és matematikai ismeretek szükségesek.
\thispagestyle{empty}
\clearpage

\noindent {\bf \large Satement of Originality}
\vspace{0.3 cm}

\noindent 
This document is written by László Horváth who declares to take full responsibility for the  contents of this document.
I declare that the text and the work presented in this document is original and that no sources other than those mentioned in the text and its references have been used in creating it.

\vspace{2 cm}

\noindent {\bf \large Önállósági nyilatkozat}
\vspace{0.3 cm}

\noindent Alulírott Horváth László, a Budapesti Műszaki és Gazdaságtudományi Egyetem fizika MSc szakos hallgatója kijelentem, hogy ezt a szakdolgozatot meg nem engedett segédeszközök nélkül, önállóan, a témavezető irányításával készítettem, és csak a megadott forrásokat használtam fel. 
Minden olyan részt, melyet szó szerint, vagy azonos értelemben, de átfogalmazva más forrásból vettem, a forrás megadásával jelöltem.
\vspace{3 cm}

\hspace{6.85 cm} \makebox[5 cm]{\dotfill}

\hspace{8 cm} Horváth László
\thispagestyle{empty}
\clearpage

\clearpage
\setcounter{page}{6}

\begin{spacing}{0.95}
\tableofcontents
\thispagestyle{empty}
\end{spacing}
\clearpage

\chapter{Introduction}\label{sec:introduction}

Our world provides plenty of options to meet the energy needs of mankind.
One of them, which is already in the service of humanity, is the release of the binding energy of atomic nuclei.
Energy is produced by nuclear fission in nuclear power plants all over the world.
However, sustainable energy production to the electricity grid by fusion reactions has not yet been achieved~\cite{eps07energy}.

\section{Fusion power generation}

An example of reliable fusion energy production is the Sun.
Its hydrogen (${}^1$H) fuel is converted to helium (${}^4$He) in different astrophysical reaction chains~\cite{bethe39energy}.
These fusion reactions provided energy for several billions of years to life on Earth.
In stellar cores the rate of fusion reactions is very slow, because proton-neutron decays have to occur.
Therefore, astrophysical reaction chains have very low cross-sections, thus the released power density of the Sun is less than $1$ W/m${}^3$~\cite{wesson99science}.

\begin{figure}[htb!]\centering
  \includegraphics[width = 0.6\textwidth]{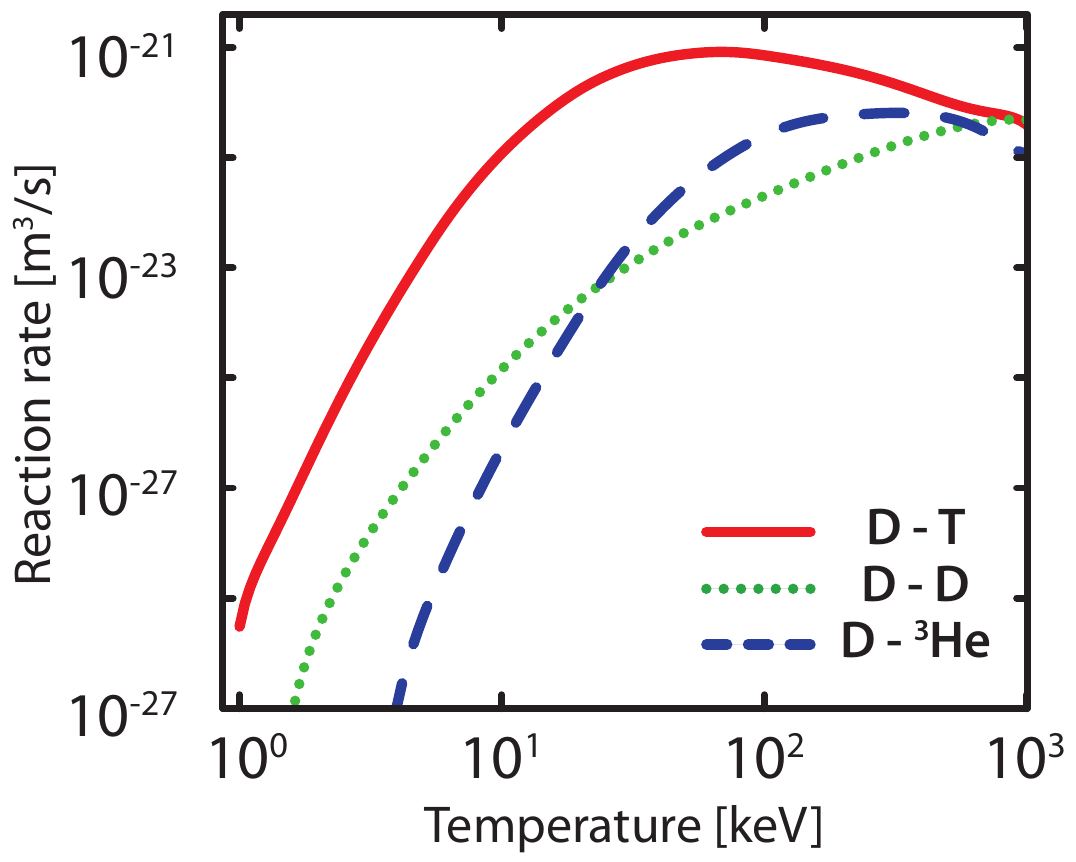}
  \caption{The fusion reaction rates of different isotopes~\cite{wesson04tokamaks}.}
  \label{fig:cross_section}
\end{figure}

Reproduction of stellar core conditions on Earth for nuclear fusion power production would be inefficient.
In order to achieve sufficiently high power from fusion reactions for electricity production, light isotopes should be chosen~\cite{wesson99science} such as deuterium (${}^2$H $\equiv$ D), tritium (${}^3$H $\equiv$ T) or ${}^3$He.
Some of the fusion reactions possible with the aforementioned isotopes are listed here~\cite{freidberg08plasma}:
\begin{alignat}{5}
  \textrm{D} & + \textrm{D}\ & \Rightarrow &\ \ {}^3\textrm{He}\ (0.82 \textrm{ MeV})\ & +\ \textrm{n}\ (2.45 \textrm{ MeV}) ,	\nonumber\\
  \textrm{D} & + \textrm{D}\ & \Rightarrow &\ \ \textrm{T}\ (1.01 \textrm{ MeV})\ & +\ \textrm{p}\ (3.02 \textrm{ MeV}) ,	\nonumber\\
  \textbf{D} & \pmb{+} \textbf{T}\ & \pmb{\Rightarrow} &\ \ {}^{\textbf{4}}\textbf{He}\ (3.52 \textrm{ MeV})\ & \pmb{+}\ \textbf{n}\ (14.1 \textrm{ MeV}) ,	\nonumber\\
  \textrm{D} & + {}^3\textrm{He}\ & \Rightarrow &\ \ {}^4\textrm{He}\ (3.66 \textrm{ MeV})\ & +\ \textrm{p}\ (14.6 \textrm{ MeV}) . \nonumber
\end{alignat}
Considering the reaction rates presented in figure~\ref{fig:cross_section}, the most favourable reaction on Earth is the deuterium-tritium (DT) reaction~\cite{dolan00fusion}.
The reaction rate of DT peaks at a lower temperature and at a higher value than other reactions commonly considered for fusion energy.

A huge amount of deuterium is available in nature, because there is 1 atom of deuterium for every 6700 atoms of hydrogen in natural water resources~\cite{freidberg08plasma}.
In addition to the enormous resources, deuterium can be easily extracted at a low cost.
However, gathering tritium is difficult.
Tritium is a radioactive isotope and it has a half-life of $12.3$ years.
Therefore, tritium can be found in nature only in trace quantities~\cite{iaea12fusion}.
Given that there is no practical tritium breeding technology which can satisfy the demand of an energy producing fusion reactor, the required tritium must be produced by the reactor itself.

\begin{figure}[htb!]\centering
  \includegraphics[width = 0.6\textwidth]{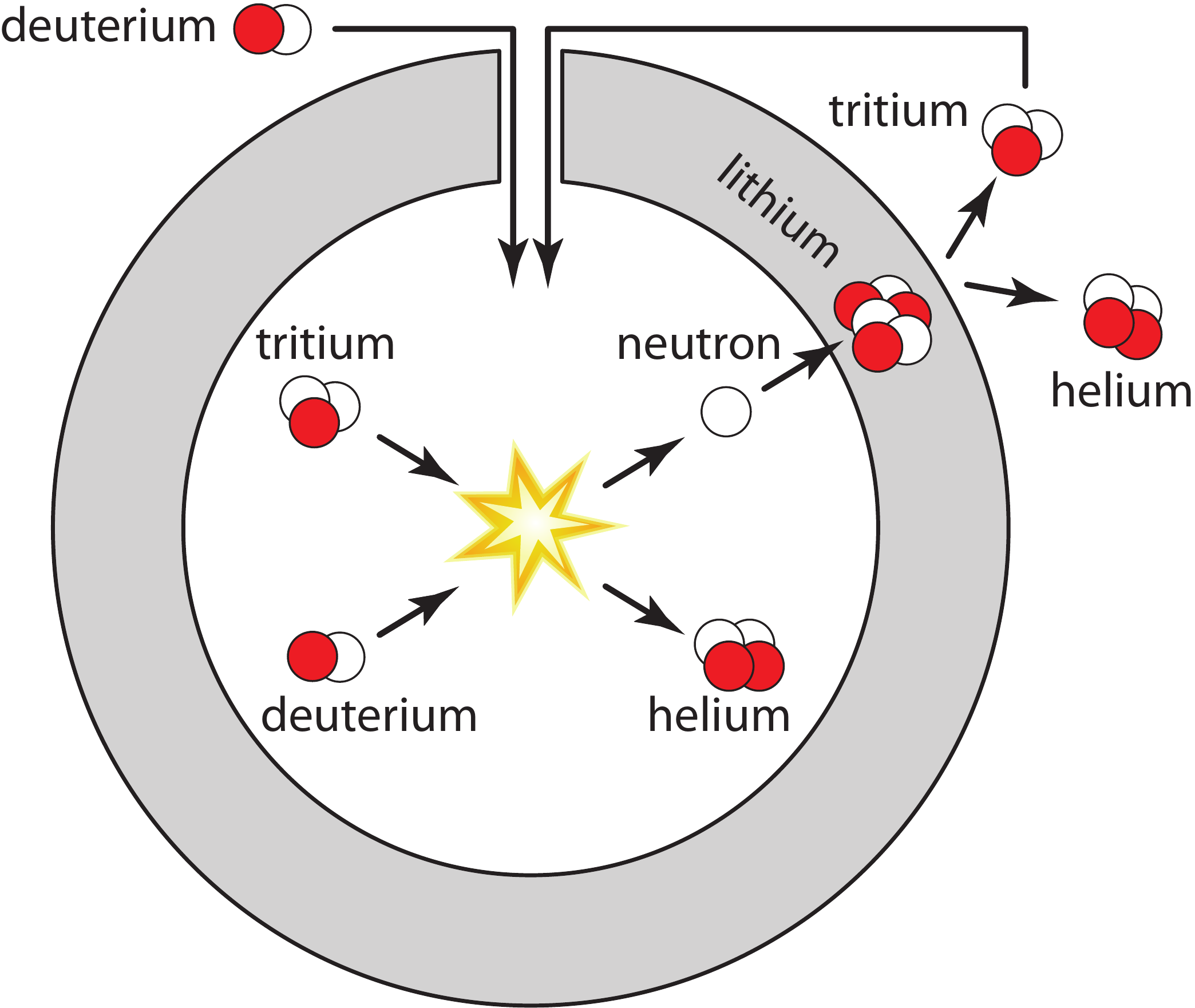}
  \caption{Operational scheme of tritium breeding. The fusion-born neutrons, by hitting the wall containing lithium, produce tritium.}
  \label{fig:tritium_breeding}
\end{figure}

Taking advantage of the fusion reaction generated neutrons, tritium can be produced from lithium using the following processes~\cite{freidberg08plasma}:
\begin{alignat}{5}
  {}^6\textrm{Li} & + \textrm{n~(thermal)}\ & = &\ \ {}^4\textrm{He}\ & +\ \textrm{T}	\nonumber\\
  {}^7\textrm{Li} & + \textrm{n~(energetic)}\ & = &\ \ {}^4\textrm{He}\ & +\ \textrm{T} & + \textrm{n} \nonumber
\end{alignat}
The tritium breeding process is illustrated in figure~\ref{fig:tritium_breeding}.
The fusion-born neutrons, by hitting the wall containing lithium, produce tritium.
In this way the fuel of a fusion power plant is deuterium and lithium.
Both are non-radioactive elements and are available in large quantities on Earth.
The final product of the fusion reaction and the tritium breeding reaction are helium, thus neither the fuel, nor the end-product of the fusion and breeding cycle is radioactive.

Fusion experiments with particle accelerators provided useful information to determine the cross sections of the aforementioned reactions.
However, sustainable energy production with accelerators cannot be achieved~\cite{wesson04tokamaks}, since the cross-section of Coulomb scattering is several orders of magnitude higher than the cross-section of the fusion reaction.
Therefore, the most promising method is to heat the DT fuel to sufficiently high temperature that the thermal velocities are high enough to produce fusion reactions.
In this case the Coulomb collisions just redistribute the energy between particles.
The necessary temperature is in the order of hundred million Kelvin for DT fuel.
The conditions required for a self-sustaining process are estimated by the Lawson criterion for DT fuel and optimum temperature ($\sim$25 keV)~\cite{wesson04tokamaks}:
\begin{equation}
  n\tau_E > 10^{20} \textrm{s/m}^3 ,
\end{equation}
where $n$ is the density and $\tau_E$ is the energy confinement time which measures the rate at which a system loses energy to its environment:
\begin{equation}
  \tau_E = \frac{W}{P_\mathrm{loss}} ,
\end{equation}
where $W$ is the plasma energy content and $P_\mathrm{loss}$ is the total power loss of the plasma.

\section{Magnetic confinement fusion}

\begin{figure}[htb!]\centering
  \includegraphics[width = 0.55\textwidth]{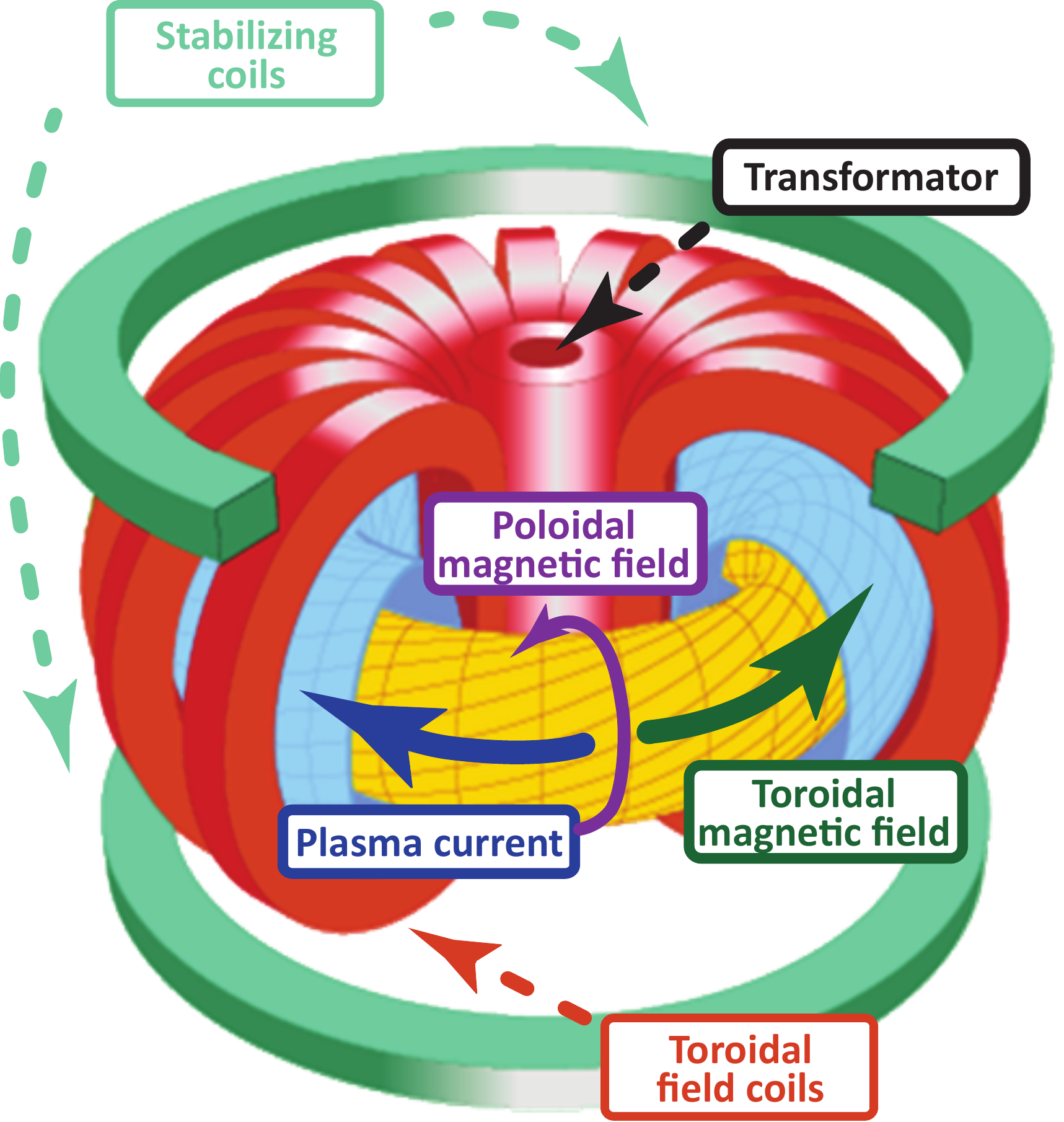}
  \caption{Schematic drawing of a tokamak. The superposition of the toroidal magnetic field and the poloidal magnetic field generated by the toroidal plasma current leads to helically twisted magnetic field lines in the torus.}
  \label{fig:tokamak}
\end{figure}

One way of confining the fuel which has such high temperature is to employ a magnetic field.
At temperatures necessary for the fusion reaction, the particles are ionized and the fuel is in the plasma state.
The charged particles (ions and electrons) start their helical motion in the plane along the magnetic field.
The circular part of this motion in the plane perpendicular to the field line is called gyromotion.
The helical motion is suitable to confine the plasma, because it does not allow the particles to move perpendicular to the magnetic field, but they are free to move parallel to the magnetic field.
The problem of losses at the ends of a linear device can be solved by bending the magnetic field lines into a torus.
However, in a toroidal geometry, the magnetic field is not homogeneous.
The strength of the magnetic field is inversely proportional to the distance measured from the centre of the torus, because of the higher current density in the toroidal field coils (indicated with red in figure~\ref{fig:tokamak}) in the inner side of the torus results in a stronger magnetic field there.
This inhomogeneity leads to a charge dependent drift ($\nabla B$ drift)~\cite{wesson04tokamaks}, which causes the electrons and ions to move in vertically opposite directions.
The resulting vertical electric field creates a charge independent $\textbf{E}\times\textbf{B}$ drift~\cite{wesson04tokamaks}, which moves the entire plasma towards the outside of the torus.
This electric field between the top and the bottom of the torus can be shorted out by helically twisting the magnetic field lines.

A type of magnetic confinement device is the tokamak~\cite{wesson04tokamaks}, where the magnetic field lines are helically twisted by a plasma current as it is illustrated in figure~\ref{fig:tokamak}.
In tokamak geometry the toroidal direction is the long way, the poloidal direction is the short way around the torus.
The toroidal plasma current is driven by a transformer coil and it twists the magnetic field lines because it generates a poloidal magnetic field indicated with purple in figure~\ref{fig:tokamak}.
The twist of magnetic field lines is described by the safety factor.
The safety factor $q$ is the ratio of toroidal transits per single poloidal transit of a magnetic field line~\cite{bellan08fundamentals}.
Due to the axisymmetry of a tokamak device, the magnetic field lines are organized into magnetic flux surfaces.
Along the magnetic field lines, the particles can move freely, thus the plasma parameters on the magnetic flux surfaces are balanced very quickly, while the transport perpendicular to the surfaces is several orders of magnitude slower.

In order to realize a self-sustaining magnetically confined plasma, the released fusion energy has to heat the plasma.
The energy generated in a fusion reaction is distributed as kinetic energy to the fusion products.
According to their mass ratio, 80 \% of the energy is carried by the neutrons which do not interact with the plasma and hit the plasma facing components.
The remaining 20 \% energy is carried by the He nuclei.
These He$^{2+}$ ions have $3.5$ MeV initial energy and isotropic angular distribution.
The self-heating provided by the He$^{2+}$ ions ($\alpha$-particles) is called $\alpha$-heating.
In a hot plasma the typical mean free path of fusion-born $\alpha$-particles is very long compared to the size of the device~\cite{lauber13super}.
Thus, it is crucial to well confine the high energy $\alpha$-particles.

Beside the $\alpha$-heating, auxiliary heating systems are needed to reach fusion relevant conditions.
Furthermore, these external heating methods are planned to be used to tailor the plasma profiles during the fusion burn.
Two types of auxiliary heating systems are used to generate fast ions, namely the neutral beam injection (NBI) and the ion cyclotron resonance heating (ICRH).
These techniques are capable of accelerating hydrogen isotopes up to the MeV energy range.
Since plasma heating scenarios are predominantly based on fast ions, their transport properties are fundamental to the success of any burning plasma experiment.
Enhanced radial transport of energetic particles (EPs) degrades the heating efficiency and it can also cause harmful power loads upon the first wall of the device.
Since, in the plasma core, the most important transport process of EPs is their interaction with global plasma waves~\cite{lauber13super}, the thorough understanding of the EP driven instabilities is crucial.

\section{Fast ion transport due to plasma waves}

Marginally stable plasma modes can be destabilized by EPs in tokamaks.
In addition to normal modes, there are also energetic particle modes (EPMs) which are forced oscillations characterized by strong dependence on the fast ion distribution function.
First, a basic introduction of the shear Alfvén waves is presented.

\begin{figure}[htb!]\centering
  \includegraphics[width = 0.6\textwidth]{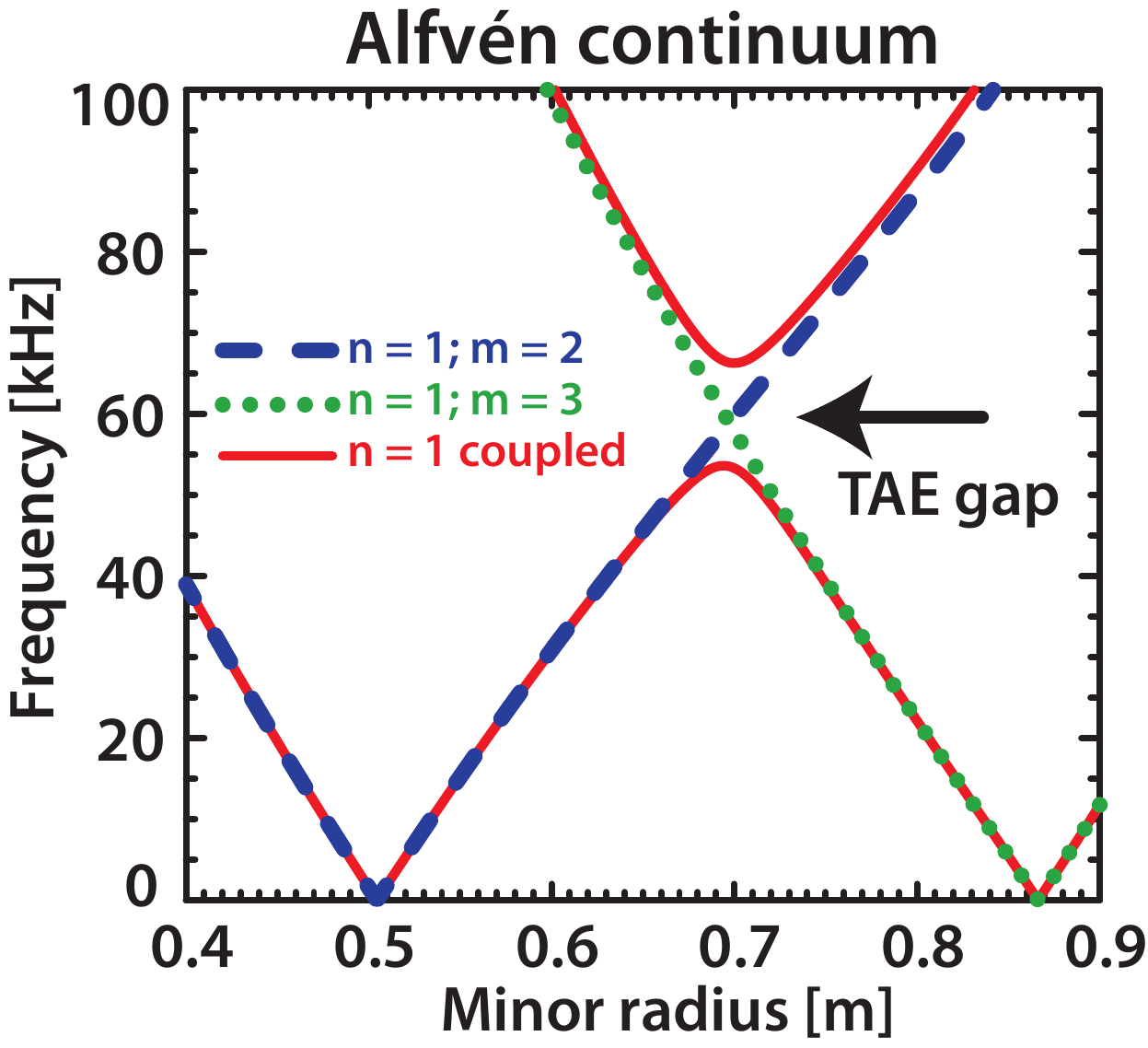}
  \caption{
  Shear Alfvén continuum plotted (without coupling) with dashed blue line (n = 1, m = 2) and with dotted green line (n = 2, m = 3). In toroidal geometry, a frequency gap is formed due to the toroidal coupling between m = 2 and m = 3 modes. The Alfvén continuum with coupling is plotted with solid red lines. In order to calculate these dispersion relations, the background plasma parameters are adopted from Lesur's PhD thesis~\cite{lesur11berk}. The major radius is 3.3 m, the minor radius is 0.96 m and the magnetic field is 1.2 T on axis. The q profile has been modeled by $q(r) = 1.2 + 2.1(r/a)^{1.5}$ and the electron density by $n_e(r) = 0.11 + 1.57 (1-r^2/a^2)^{0.3}$ [$10^{19}$ m$^{-3}$].}
  \label{fig:tae_dispersion}
\end{figure}

Shear Alfvén waves are electromagnetic waves that propagate along the magnetic field with the following dispersion relation~\cite{freidberg1987ideal}:
\begin{equation}\label{eq:dispersion1}
  \omega^2 = k^2_\parallel v^2_A ,
\end{equation}
where
\begin{equation}
  v_A = \frac{B}{\sqrt{\mu_0\sum_i n_i m_i}}
\end{equation}
is the Alfvén velocity, $k_\parallel$ is the wave number in the direction of the magnetic field, $\sum_i n_i m_i$ is the mass density of the plasma, $i$ denotes the different ion species, $B$ is the magnitude of the magnetic field and $\mu_0$ is the vacuum permeability.
In a torus, the periodic boundary condition requires that the wave solutions are quantized in toroidal and poloidal directions:
\begin{equation}
  k_\parallel = \frac{n-m/q(r)}{R_0} ,
\end{equation}
where $n$ and $m$ are the toroidal and poloidal mode numbers respectively, $q$ is the safety factor and $R_0$ is the major radius of the plasma.
In a toroidal plasma both $k_\parallel$ and $v_A$ are functions of the minor radius $r$.
This leads to the so called shear Alfvén continuum.
The dispersion relation of two successive modes with poloidal mode number $m = 2$ and $3$ is illustrated on figure~\ref{fig:tae_dispersion}.
Waves at different radii have different phase velocities which leads to phase-mixing and a hypothetical wave packet would rapidly disperse.
This process is called continuum damping~\cite{lesur11berk}.
In a torus with a large aspect ratio ($R_0 \gg a$) the magnetic field can be approximated by $B\approx~B_0(1-\epsilon\cos(\theta))$~\cite{wesson04tokamaks}, where $\epsilon~=~a/R_0$ is the inverse aspect ratio and $\theta$ is the poloidal coordinate.
The parallel component of the wave vector now also depends on the poloidal coordinate $\theta$ which can lead to intersection of the dispersion relations belonging to different poloidal mode numbers $m$.
The crossing is expected to be at the radius where the wave number of the two counter-propagating waves equals:
\begin{equation}
  k_{m_1}(r_m) = -k_{m_2}(r_m) .
\end{equation}
The crossing of $m = 2$ curve (blue) and the $m = 3$ curve (green) is visible in figure~\ref{fig:tae_dispersion}.
However, toroidicity resolves this degeneracy at the points of intersection and produces gaps in the frequency continuum as shown in figure~\ref{fig:tae_dispersion} with red.
For a particular toroidal mode number, and to the lowest order in the inverse aspect ratio ($\epsilon=a/R_0$) expansion parameter, each poloidal harmonic couples to its nearest neighbouring sidebands~\cite{pinches96nonlinear} and the eq.~\eqref{eq:dispersion1} dispersion relations changes:
\begin{equation}\label{eq:dispersion2}
  \omega^2_{1,2}/v^2_A = \frac{k^2_{\parallel m} + k^2_{\parallel m+1} \pm
    \sqrt{(k^2_{\parallel m} - k^2_{\parallel m+1})^2 + 4\epsilon^2 k^2_{\parallel m} k^2_{\parallel m+1}}}
    {2(1-\epsilon)} .
\end{equation}
Within these gaps global modes can exist, because waves do not experience heavy continuum damping.
These modes are called toroidicity-induced Alfvén eigenmodes (TAEs).
Analogous to the TAE gap further gaps arise caused by the coupling of the $m$, $m+2$ and $m$, $m+3$ modes such as elipticity induced gap and non-up-down-symmetric gap~\cite{lauber13super}.
In these gaps radially extended, weakly damped modes can exist, which can be driven unstable by the fast ion population.
The properties of these eigenmodes such as frequency or mode structure are mostly determined by the background plasma.
A similar gap mode is the $\beta$-induced Alfvén eigenmode (BAE).
The BAE gap is introduced by the coupling between compressible acoustic waves and the shear Alfvén continuum~\cite{turnbull93global}.

These modes can be driven unstable by the free energy in the fast ion distribution function~\cite{lauber13super}.
Energy transfer between an ion and a wave requires that the velocity vector is not perpendicular to the electric field of the wave.
Thus, only the drift velocity of ions contributes to the energy transfer~\cite{heidbrink08basic}.
The energy transfer associated with the gyromotion averages to zero, since the gyromotion is very rapid compared to the mode frequency.

The growth rate of a mode is generally small compared to the wave frequency, because significant net energy transfer is gained only after dozens of orbital cycles.
To avoid phase averaging to zero, a harmonic of the drift-orbit frequency must match the wave frequency, so the following resonance condition has to be fulfilled for net energy gain~\cite{heidbrink08basic}:
\begin{equation}\label{eq:resonance_condition}
  \omega + (m + l)\omega_{\theta} - n\omega_{\xi} \simeq 0 ,
\end{equation}
where $\omega$ is the wave frequency, $\omega_{\theta}$ is the frequency of the poloidal motion of the ion, $\omega_{\xi}$ is the frequency of the toroidal motion of the ion, $m$ is the poloidal mode number and $n$ is the toroidal mode number of the mode.
$l$ denotes the order of the harmonic of the drift-orbit frequency.
In strongly shaped plasmas and for large drift-orbit displacements, higher order harmonics are also important.

The net energy exchange between the wave and the particles is determined by several factors.
It depends on the amplitude of the particular poloidal harmonic and on the drift velocity.
Furthermore, the magnitude of the energy exchange also depends on the gradient associated with the particular resonance in the EP distribution function.
Particles with speed slightly below the resonant velocity gain energy, while particles with speed slightly above the resonant velocity lose energy.
Thus, EPs can drive a mode where the gradient in the EP distribution is positive along the coordinate defining the resonance.
Mostly the gradient in the EP distribution respect to the energy is negative, so the energy distribution usually damps the wave.
This is true even for the slowing-down distribution in a burning plasma.
Modes are usually driven by the free energy in the spatial gradient.
However, the anisotropy in the velocity space can also drive the mode.
Finally, the alignment of the particle orbit and the eigenmode is also a governing factor of the energy exchange.
Modes are excited if the drive can overcome the damping of the background plasma.
Important damping mechanisms include ion Landau damping, continuum damping and radiative damping~\cite{lauber13super}.
The damping rates are very sensitive to plasma parameters such as the temperature and the q profile.

Several types of non-linear behaviour of the mode amplitude and frequency are observed on present-day tokamaks.
This behaviour ranges from steady-state -- where the amplitude saturates at nearly constant frequency -- to an explosive growth.
The type of behaviour is expected to significantly influence the impact of the instabilities on the fast particle transport.
Thus, their thorough understanding from both experimental and theoretical side is essential.
In this thesis modes with bursting amplitude and rapidly changing mode frequency are experimentally investigated in detail.
These modes are also called ``chirping'' modes.

So far, the discussion was restricted to Alfvén modes, but if the EP pressure is sufficiently high compared to the thermal pressure, the EP drive can overcome the continuum damping and energetic particle modes are excited.
Their mode structure is independent from the shear Alfvén continuum, and sensitively depends on the EP distribution.
Their frequency usually correspond to a characteristic frequency of the EP orbital motion.

One of the main open questions concerning EP driven Alfvén eigenmodes and EPMs is the non-linear evolution of the mode amplitude.
These instabilities constitute a non-linear system where kinetic and MHD non-linearities can both be important making it difficult to describe the phenomenon.
In this thesis I present my results on the experimental investigation of beta induced Alfv\'{e}n eigenmodes (BAEs) and EP-driven geodesic acoustic modes (EGAMs) observed in the ramp-up phase of off-axis NBI heated plasmas in ASDEX Upgrade (AUG)~\cite{lauber13offaxis}.

The plasma scenario in which these modes were observed needs low density operation and off-axis NBI heating.
Low density is important in order to keep the ratio of fast ion pressure and thermal pressure as high as possible.
The excitation of these modes is possible if the damping is low enough.
The damping of EGAMs is lower at lower temperatures.
Thus, they are easily driven unstable at lower temperatures e.g. during the ramp-up phase of a discharge.
The main damping mechanism of BAE is the ion Landau damping~\cite{nguyen09excitation} which scales with the background ion temperature and density.
This way in low density, low temperature plasmas BAEs are more easily excited.
On AUG, even in low density operation, the damping of BAEs in the flat-top -- where the main plasma parameters are maintained at constant values -- is not sufficiently small to excite them with NBI.
However, in the ramp-up phase BAEs are routinely driven unstable with off-axis NBI heating.
The role of off-axis NBI is twofold.
One is that, further out from the magnetic axis, the temperature and density is lower which again allows the excitation of these modes.
Second, the dominant auxiliary heating system in future burning plasma experiments will be off-axis NBI.

The confinement of fast ions is crucial in a burning plasma experiment such as ITER where the heating of the plasma is based on the dominant self-heating by fusion-born $\alpha$-particles.
ITER is a large-scale scientific experiment under construction that aims to demonstrate the technological and scientific feasibility of fusion energy~\cite{ikeda07progress}.
In general, EP driven modes have a massive contribution to the radial transport of fast ions.
However, in view of that bursting BAEs and EGAMs are only observed in the ramp-up phase of discharges where damping is low enough due to the low density and temperature the question arises what are their implications regarding ITER or any other burning plasma experiment?

BAEs are driven by the radial gradient in the EP distribution function~\cite{nguyen09excitation}.
Thus, their presence enhances the radial transport of fast ions.
However, their stability in ITER is not yet clear.
Simulations have shown that in the ITER inductive baseline scenario the radial gradients in the EP distribution will not be sufficiently high to destabilize BAEs, but the difference is marginal and the uncertainty of these results is high~\cite{pinches15energetic}.
EGAMs themselves do not cause radial redistribution of the fast ions directly, because they are driven by the velocity phase space gradient in the EP distribution.
Therefore, the isotropic distribution of fusion-born $\alpha$ particles cannot affect EGAMs.
However, NBI driven EGAMs can influence fast ion losses indirectly via mode-mode coupling with TAEs~\cite{lauber15personal}.
The impact of BAEs and EGAMs on the fast ion confinement in ITER is not completely clear.
On the other hand, the investigation of these modes is essential to understand the non-linear behaviour of EP driven modes.

The main goal of this thesis is to experimentally investigate the rapid changes in the radial structure of bursting EP-driven modes during the non-linear chirping phase.
Due to diagnostic and data analysis complexities this task has never been accomplished before.
Some modes are expected to retain their radial structure, while others would be expected to change.
Even qualitative results can provide important information about the underlying physics and strengthen (or challenge) our present theoretical understanding.
Furthermore, an analysis based on the methods developed here can serve as the basis of comparison with numerical codes which simulate the time evolution of EP-driven modes.

Typically, plasma modes cause magnetic, density and temperature fluctuations in the plasma which makes it possible to extensively examine them in a well-equipped tokamak such as AUG.
Since the observed modes cause weak oscillations in the measured quantities, the used diagnostics tools were carefully selected by taking into consideration their spatial and temporal resolution and their signal-to-noise ratio.
The argument of the selection and the description of the applied diagnostic tools are presented in section~\ref{sec:measurement_setup}.
The analysis in question requires sophisticated signal processing tools.
Since this work is focused on the investigation of transient, wave-like phenomena, the signal processing methods developed and used are based on short time Fourier transform (STFT)~\cite{gabor1946analysis}.
The applied methods are described in section~\ref{sec:analysis_principles}.
The amplitude reconstruction method derived in this thesis which allowed to investigate the radial structure of the observed modes is presented in chapter~\ref{sec:amp_reconstruction}.
The dedicated experiments analysed are shown in section~\ref{sec:results} where the results are also discussed in view of the physical picture.

\chapter{Measurement set-up and analysis principles}\label{sec:setup}

The frequency of the modes investigated in this thesis is in the order of $50$~-~$100$~kHz.
Both the diagnostic tools and the data processing method were chosen to deal with such high frequency oscillations.

From the diagnostic point of view, fluctuation measurements are required which can measure either the magnetic, the density or the temperature fluctuations in the plasma with higher temporal resolution than $100$~kHz.
The possible candidates were the magnetic pick-up coil, the reflectrometry, the electron cyclotron emission (ECE) and the soft X-ray (SXR) measurements.
Magnetic measurements do not have spatial resolution, thus it is not suitable for the analysis of the radial structure, but it was used to examine the mode structure.
The reflectrometry diagnostic can measure density fluctuation with high temporal resolution, however due to the low spatial resolution I only used it to localize the observed modes.
The ECE~\cite{hutchinson02principles} and the ECE imaging (ECEI)~\cite{classen10electron} diagnostics provide local electron temperature measurements.
These would be good candidates due to their high time- and spatial resolution, however BAEs and EGAMs were not visible by ECE and ECEI in the investigated shots.
Thus, ECE and ECEI are ruled out from the analysis.
The signal of SXR measurement is a function of electron temperature and density.
Its sampling frequency is $2$ MHz and due to the high number of SXR channels it also has a spatial resolution.
Therefore, this diagnostic was chosen to investigate the changes in the radial structure of the modes.
The magnetic probes, the soft X-ray measurements and the reflectrometry diagnostics are introduced in section~\ref{sec:measurement_setup}.

In addition to the suitable diagnostic tools, an appropriate data processing method is required.
For this purpose, continuous linear time-frequency transforms were chosen, because these are ideal to investigate transient signals.
The amplitude reconstruction method developed is presented later in chapter~\ref{sec:amp_reconstruction} is based on STFT~\cite{gabor1946analysis} which is a linear time-frequency transform.
The mathematical background of these transforms are introduced in section~\ref{sec:analysis_principles}.

\section{Measurement setup}\label{sec:measurement_setup}

The experiments presented in this thesis were carried out on the ASDEX Upgrade tokamak (Axially Symmetric Divertor EXperiment Upgrade)~\cite{stroth13overview}, which is a middle sized tokamak with a major radius of $1.6$ m and a minor radius of $0.5$ m.
It started operation at the Max-Planck-Institut für Plasmaphysik, Garching in 1991 and it plays a substantial role in the European fusion program~\cite{efda13roadmap}.
At present, it is Germany's largest fusion experiment.
The device is equipped with a flexible neutral beam injection (NBI) system which allows to carry out cutting edge research in fast ion physics.
Furthermore, the tokamak has a wide variety of diagnostics tools.
Three of them were used in this thesis and these are presented in the forthcoming sections.

\subsection{Magnetic probes}\label{sec:magnetic}

The magnetic pick-up coils are mounted on the inner side of the vacuum chamber.
Depending on the alignment these probes pick up different components of the magnetic field fluctuation.
On AUG, so-called Mirnov coils are placed on the vacuum vessel to measure the poloidal component of the magnetic field fluctuations $\tilde{B}_\mathrm{pol}$ and so-called ballooning coils are placed closer to the plasma on the low field side to measure the radial component $\tilde{B}_\mathrm{r}$.
Due to their relative simplicity and compact size, several coils can be distributed over different locations within the plasma vessel, making them suitable for the analysis of mode structure.

\begin{figure}[htb!]\centering
  \includegraphics[width = 1.0\textwidth]{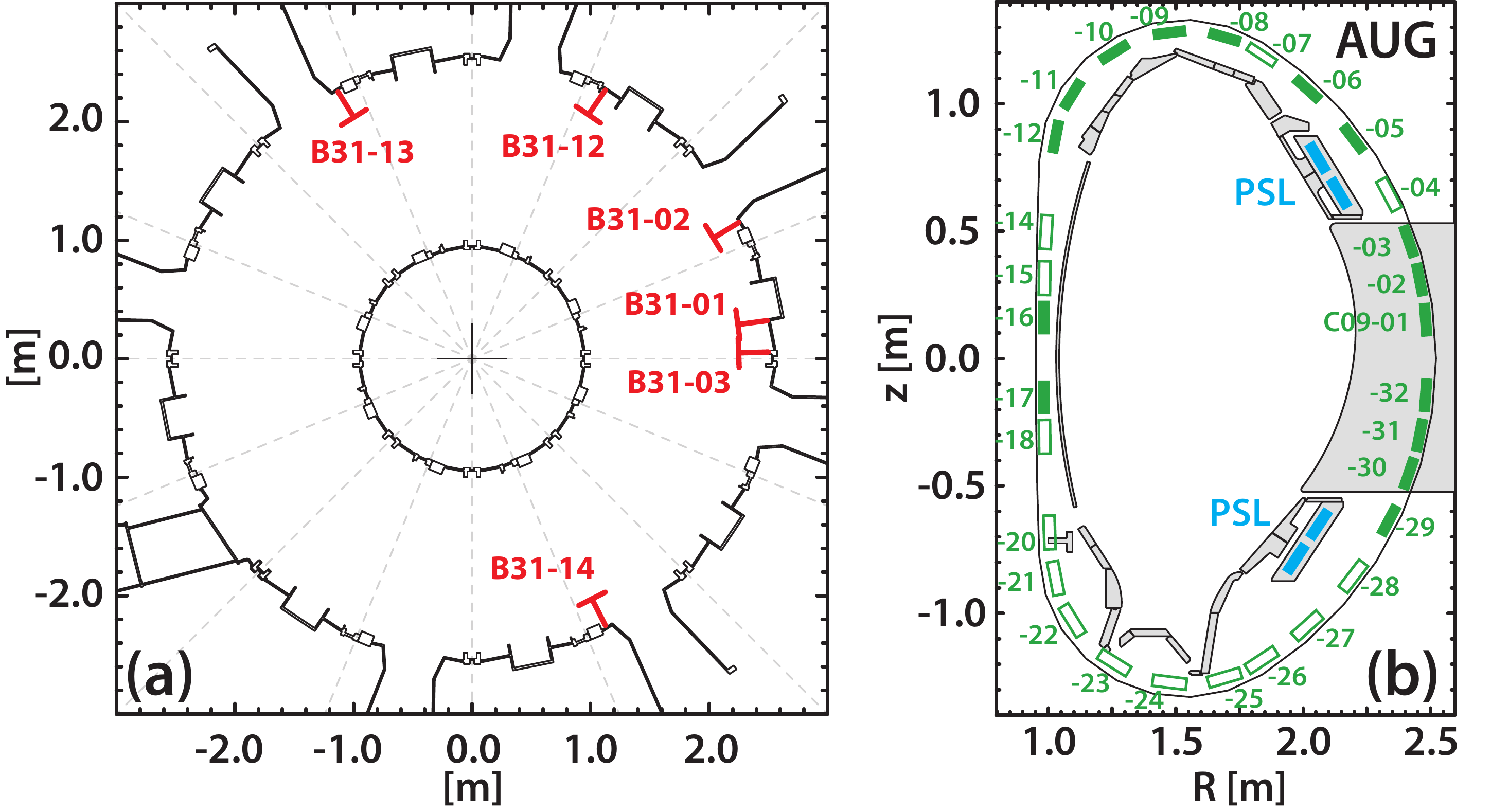}
  \caption{The position of magnetic probes used for the analysis on AUG. \textbf{(a)}~The six ballooning coils of the toroidal array are indicated with red in the top-down view of the tokamak. \textbf{(b)}~The 30 Mirnov coils indicated with green in a poloidal cross section of the device. Probes indicated with empty rectangles were ruled out from the analysis, because the investigated modes were not visible in their signal. The cyan rectangles indicate the passive stabilizing loops (PSLs).}
  \label{fig:magnetic_probe_position}
\end{figure}

The toroidal position of the ballooning coils is illustrated in figure~\ref{fig:magnetic_probe_position}a where a top-down view of the tokamak is shown.
These probes are placed in almost identical poloidal positions which allows to investigate toroidal mode numbers.
Furthermore, these coils are placed near the last closed flux surface to more efficiently detect the perturbations coming from the plasma.
The ballooning coils are calibrated by taking into account the transfer functions of the different probes~\cite{horvath15reducing}.
This way the systematic errors in the mode number fitting are reduced which allows a better identification of mode numbers with reduced number of coils even for transient modes.

In order to determine the poloidal mode number, the poloidal array of Mirnov coils was used.
Mirnov coils are placed on the vacuum vessel, a little further from the plasma and these are not calibrated.
In turn, the 30 Mirnov-coils of the poloidal array constitute a full poloidal ring around the plasma.
The position of these probes is shown in figure~\ref{fig:magnetic_probe_position}b in a poloidal cross section of AUG.
Probes indicated with empty green rectangles are ruled out from the mode number analysis because the observed modes were hardly seen in their signals.
Most of the ruled out coils have a weak signal because these are placed behind the divertor elements or the passive stabilizing loops (PSLs).
Despite the difficulties concerning the use of Mirnov-coils, due to the high number of available probes, the poloidal mode number analysis gives accurate results for low mode numbers.

\subsection{Soft X-ray measurement}\label{sec:sxr}

The magnetic probes are very useful in mode identification and mode number analysis, but they cannot provide information on the radial structure of plasma modes.
For this purpose, an other fluctuation measurement was chosen: the soft X-ray diagnostic system.
The major part of the radiation coming from a hot fusion plasma is in the soft X-ray range (100~eV~to 10~keV).
The energy spectrum of Soft X-ray radiation consists of a continuum radiation resulting from Bremsstrahlung, a recombination radiation resulting from free-bound transition, and a line radiation resulting from bound-bound transition.
In a plasma the light intensity from Bremsstrahlung radiation can be estimated by the following formula~\cite{huba09nrl}:
\begin{equation}\label{eq:sxr_intensity}
  P_{\mathrm{Br}} = 1.69\cdot10^{-32}\cdot n_e\sqrt{T_e}\sum \big( Z^2 N(Z) \big) ,
\end{equation}
where $n_e$ is the electron density in $1/\mathrm{cm}^{-3}$, $T_e$ is the electron temperature in eV and $Z$ is the charge number of the given ionization state.
The sum is executed on all ionization states and the radiation power is given in W/$\mathrm{cm}^{-3}$.
Eq.~\eqref{eq:sxr_intensity} tells us that the Bremsstrahlung emitted by a deuterium plasma is proportional to the square of the electron density and the square root of the electron temperature.
However, the soft X-ray radiation also involves recombination and line radiation.
On AUG, the SXR detectors have a 75 $\mu$m beryllium filter foil in front to suppress low energy photons.
This way, photon energies below 1 keV are blocked by the beryllium and line radiation from light impurities is negligible for the Soft X-ray diagnostic.
Thus, they affect the Soft X-ray radiation mainly via effective charge number $Z_{\mathrm{eff}}$.
On the other hand, due to the full tungsten wall coverage of AUG, line radiation can even exceed the continuum radiation seen in Soft X-ray diodes.
From the analysis point of view, it is sufficient to know that the measured radiation is a function of electron temperature and density.

\begin{figure}[htb!]\centering
  \includegraphics[width = 0.45\textwidth]{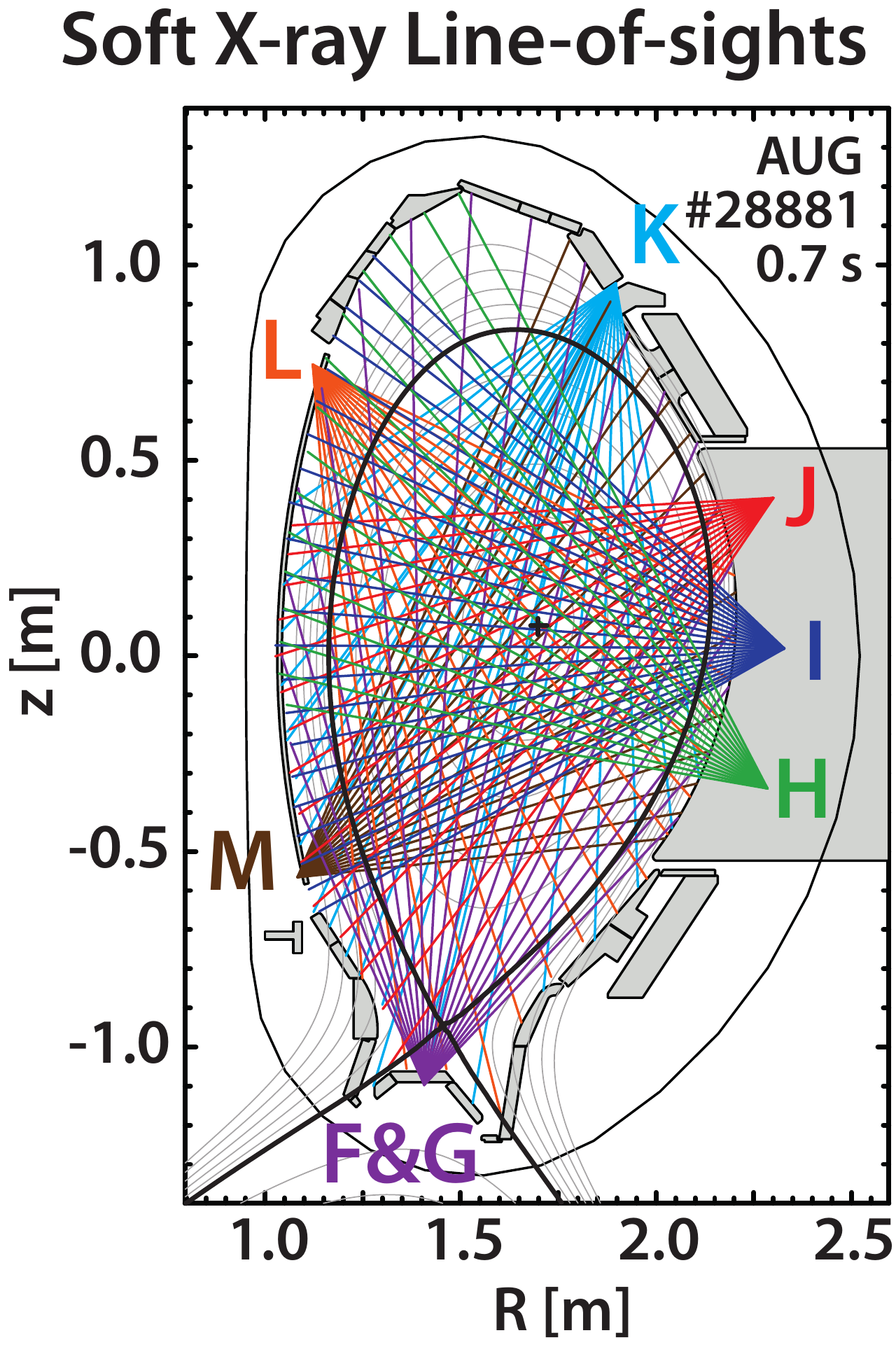}
  \caption{The line-of-sights (LOSs) of the used soft X-ray channels in the poloidal cross-section of AUG. Different colours indicate the different cameras.}
  \label{fig:sxr_position_all}
\end{figure}

The SXR system of AUG consists of planar photo diode arrays with pinhole optics.
A diode collects light from a thin cone shaped volume which is referred to as the line-of-sight (LOS).
AUG is equipped with 8 cameras consisting about 200 LOSs~\cite{igochine10hotlink}.
During the analysis of chirping modes I examined the phenomenon on the majority of the LOSs that are shown in figure~\ref{fig:sxr_position_all}.

\subsection{Reflectrometry}

Reflectrometry can give information on the density of the plasma.
When an electromagnetic wave of a certain frequency propagates through a plasma with density increasing in the direction of propagation, it may arrive at a point where the electron density equals the cut-off density~\cite{bellan08fundamentals}.
At this point the wave is reflected and by detecting the reflected wave it is possible to use it to diagnose the plasma density and its fluctuation~\cite{hutchinson02principles}.
The phase shift of the reflected wave -- which can be measured by using a microwave interferometer -- gives information on the position of the cut-off layer.

One operation mode of the diagnostic system is the profile measurement.
By sweeping the frequency of the incident wave and recording the phase shift of the reflected wave as a function of the frequency, a radial density profile can be reconstructed.
An other operation mode is the density fluctuation measurement which is carried out at a fixed frequency.
From the time variation of the phase shift, density fluctuations near the cut-off layer can be determined.
If the plasma is probed by several incident waves, the radial localization of the mode can be estimated.
On AUG there are 4 channels for fluctuation measurements both on the low field side and the high field side.

\section{Analysis principles}\label{sec:analysis_principles}

To handle the rapidly changing mode frequency and the low signal-to-noise ratio of the observed EP-driven modes, an advanced amplitude reconstruction method was developed.
This is based on Short Time Fourier Transform (STFT) which is a linear continuous time-frequency transform.
It is ideal to investigate the time evolution of transient wave-like phenomena.
The mathematical background of linear continuous time-frequency transforms is presented in section~\ref{sec:stft} based on the book of Stephane Mallat: A Wavelet Tour of Signal Processing~\cite{mallat08wavelet}.
Then, in section~\ref{sec:modenumber}, the mode number analysis method which is used to identify EP-driven modes is briefly explained.
The introduction of analysis principles continues in section~\ref{sec:amp_reconstruction}, where the amplitude reconstruction method developed in this thesis is presented.

\subsection{Short time Fourier transform}\label{sec:stft}

Linear continuous time-frequency transforms are calculated by expanding the signal $f(t)$ on the basis of families of so-called time-frequency atoms:
\begin{equation}\label{eq:transform}
  T f(u, \xi) = \langle f, g_{u, \xi} \rangle = \int\limits_{-\infty}^{+\infty} f(t) g^*_{u, \xi}(t) \dt ,
\end{equation}
where $g_{u, \xi}$ is a time-frequency atom, whose energy is well localized in both time and frequency.
Variables $u$ and $\xi$ are the time and frequency indices of the atom identifying its position on the time-frequency plane, and the $^*$ represents the complex conjugation.
The energy density distribution can then be calculated by taking the absolute value squared~\cite{mallat08wavelet}:
\begin{equation}\label{eq:energy_density}
  E f(u, \xi) = |T f(u, \xi)|^2 .
\end{equation}
$E f(u, \xi)$ can be interpreted as energy density distribution on the time-frequency plane when the signal energy is defined as
\begin{equation}\label{eq:signal_energy}
  P = \int\limits_0^T |f(t)|^2 \dt .
\end{equation}
The mode number analysis is based on the phase of the cross-transform which is defined in the following way:
\begin{equation}\label{eq:cross-phase}
  \varphi_{kl} (u, \xi) = \arg\left\{T f_k(u, \xi) T f_l^*(u, \xi)\right\} ,
\end{equation}
where $f_k$ and $f_l$ represent the signals of probes placed in different positions.

\begin{figure}[htb!]\centering
  \includegraphics[width = 0.80\textwidth]{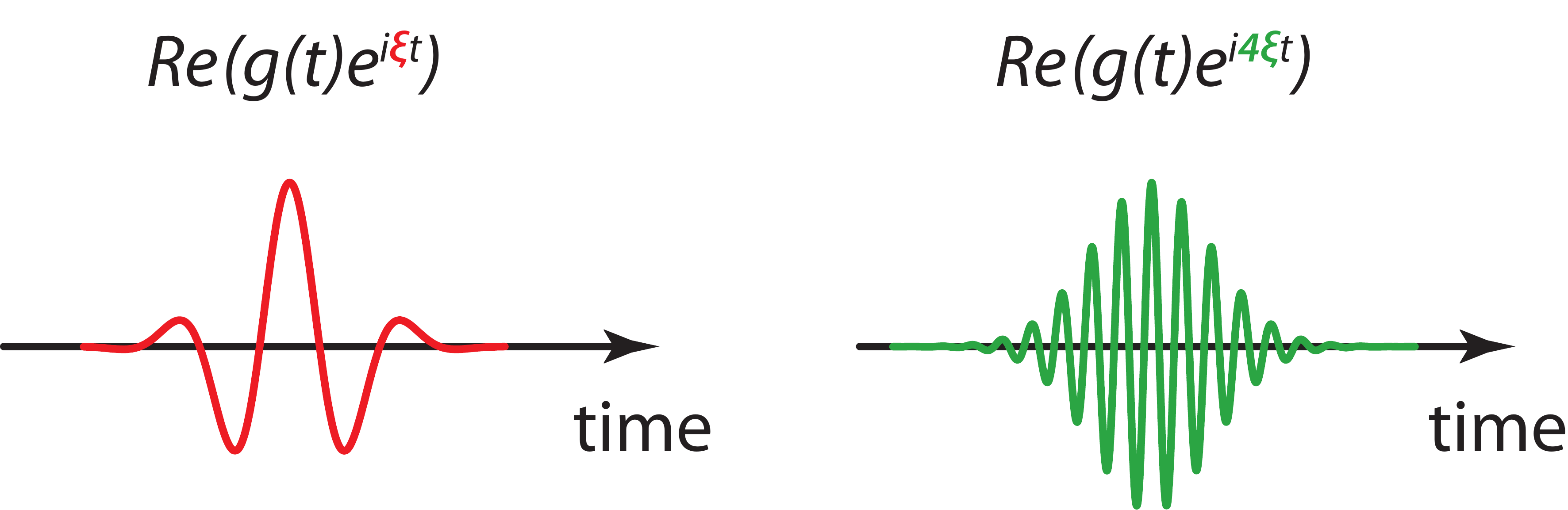}
  \caption{Two time-frequency atoms shifted in frequency~($\xi$).}
  \label{fig:atom}
\end{figure}

STFT is a type of linear continuous time-frequency transform when the family of time-frequency atoms are generated by shifting a real and symmetric window $g(t)$ in time $(u)$ and frequency $(\xi)$:
\begin{equation}\label{eq:stft-atom}
  g_{u, \xi}(t) = \exp\{i\xi t\} g(t - u) ,
\end{equation}
which gives a uniform time-frequency resolution on the time-frequency plane.
Figure~\ref{fig:atom} illustrates two atoms shifted in frequency~($\xi$).
Following from eq.~\eqref{eq:transform} STFT can be written as
\begin{equation}\label{eq:stft}
  S f(u, \xi) = \langle f, g_{u, \xi} \rangle = \int\limits_{-\infty}^{+\infty} f(t) \exp\{-i \xi t\} g(t - u) \dt .
\end{equation}
I applied Gabor atom~\cite{mallat08wavelet} in the present thesis which means that $g(t)$ is a Gaussian function with a $\sigma_t$ standard deviation:
\begin{equation}\label{eq:gabor_atom}
  g(t-u) = \frac{1}{\sqrt{\sqrt{\pi}\sigma_t}}\exp\left\{-\frac{(t-u)^2}{2\sigma_t^2}\right\} .
\end{equation}

\begin{figure}[htb!]\centering
  \includegraphics[width = 0.6\textwidth]{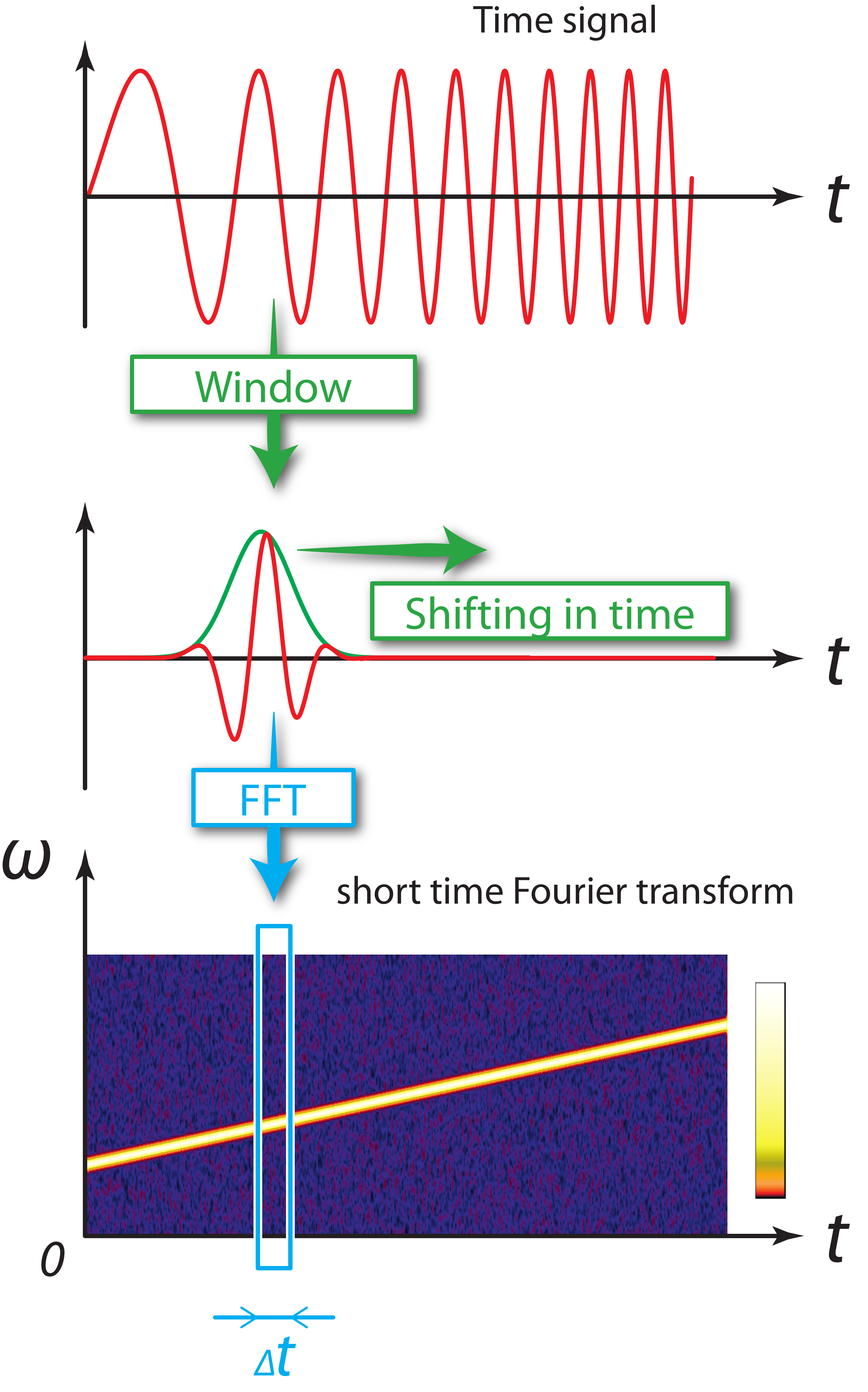}
  \caption{The process of calculating STFT on discrete time signals. The transform is calculated in each time step by fast Fourier transform (FFT) using Gaussian window.}
  \label{fig:stft_demo}
\end{figure}

The energy density distribution defined in eq.~\eqref{eq:energy_density} calculated from STFT is called a spectrogram, which is ideal to investigate the time-frequency evolution of non-stationary modes~\cite{sertoli13characterization, pokol13continuous}.
In order to apply a continuous transform on sampled time signals, the calculation has to be discretized in a way to avoid the degradation of the time-shift invariance property of the transform~\cite{pokol07application}.
This is achieved by replacing the integrals with sums and discretizing all variables in equation \eqref{eq:stft} with the smallest possible steps.
The width ($2 \sigma_t$ in equation \eqref{eq:gabor_atom}) of the Gabor atom determines the time-frequency resolution of the transform.
According to the specific application, the discretization can be done on a more sparse grid, which is determined by a parameter called time step, but it has to be at most the quarter of the width of the atom to give a continuous transform.
Since the Gabor atom has an infinite support it has to be truncated \cite{pokol07application}.
To preserve the appealing properties of the continuous transform, only the part where its values are several orders of magnitude smaller than the maximum are neglected.
The resulting length of the window determines the frequency step of the transform.
In practice the transform is calculated in each time step by fast Fourier transform (FFT) using Gaussian window having very long support as it is illustrated in figure~\ref{fig:stft_demo}.

\subsection{Mode number analysis}\label{sec:modenumber}

The evaluation of toroidal and poloidal mode numbers of the observed EP-driven modes were carried out by using signals of the magnetic pick-up coil system which is described in section~\ref{sec:magnetic}.
A brief introduction of the mode number analysis method I used is given in this section, but a more detailed description can be found in our recently submitted paper~\cite{horvath15reducing}.

The measurement of magnetic field fluctuations produced by magnetohydrodynamic (MHD) modes allows the reconstruction of their mode structure.
Harmonics of global MHD eigenmodes are generally assumed in the following form~\cite{zegenhagen06analysis}:
\begin{equation}\label{eq:eigenmode}
  A_{m,n}(\rho,\theta^*,\phi,t) = A_{m,n}(\rho,\theta^*)\exp\{i(m\theta^*+n\phi-\omega t)\} \ ,
\end{equation}
where $A$ is the observable quantity (in our case the magnetic field perturbation) as a function of time $t$ and the so-called straight field line coordinates $(\rho,\theta^*,\phi)$~\cite{schittenhelm97analysis}, $A(\rho,\theta^*)$ is the radial eigenfuntion and $\omega$ is the mode frequency.
The mode structure of the mode is characterized by the $m$ poloidal and $n$ toroidal mode numbers.
As follows from eq.~\eqref{eq:eigenmode}, in the case of a single, pure sinusoidal, globally coherent mode at fixed $(\rho,\theta^*,t)$, but at different toroidal angles, the relative phase between $A(\phi_k)$ and $A(\phi_l)$ is proportional to the relative toroidal angle $\phi_k - \phi_l$ of the probe locations and the ratio is the toroidal mode number~$n$:
\begin{equation}\label{eq:proportional1}
  \varphi_{kl}^{\mathrm{tor}}{}_{\big|_{\theta_k = \theta_l}} = 
  \arg\left\{ A(\phi_k) \right\} - \arg\left\{ A(\phi_l) \right\} = n(\phi_k - \phi_l) \ .
\end{equation}
Similar formula can be constructed for the poloidal mode number at fixed toroidal position:
\begin{equation}\label{eq:proportional2}
  \varphi_{kl}^{\mathrm{pol}}{}_{\big|_{\phi_k = \phi_l}} = 
  \arg\left\{ A(\theta_k) \right\} - \arg\left\{ A(\theta_l) \right\} = m(\theta_k - \theta_l) \ .
\end{equation}
Due to its complexity, the $\theta^*$ transformation of the poloidal coordinates was not performed.
This causes negligible error, because the investigated modes are core localized, thus the effect of the $\theta^*$ transformation would be small.

Considering eq.~\eqref{eq:proportional1} and \eqref{eq:proportional2} the mode number determination can be handled as a linear fitting problem.
In order to evaluate the toroidal mode number, one has to determine the phase of the mode in different toroidal locations $(\phi_i)$, but at the same poloidal position.
Eq.~\eqref{eq:proportional1} shows that the slope of the linear curve, fitted on relative phases between all pairs of signals as a function of the relative probe position, gives the toroidal mode number~$n$.
A similar argument stands for the poloidal mode number by considering eq.~\eqref{eq:proportional2}.
The cross-phases are evaluated from the cross-STFT which is defined in eq.~\eqref{eq:cross-phase}.
The method searches for the integer mode number value where the residual $Q$ of the fit is minimal:
\begin{equation}\label{eq:residual}
  Q = \sum_{\mathcal{P}} \| \varphi_{\mathcal{P}} - n\phi_{\mathcal{P}} \|^2 \ ,
\end{equation}
where $\varphi_\mathcal{P}$ is the phase between pairs of signals, $\phi_\mathcal{P}$ is the relative probe position, $n$ is the toroidal mode number, the sum is executed on the $\mathcal{P}$ signal pairs and $\| . \|$ is the norm obtained by taking the optimum shift of $\varphi_{\mathcal{P}}$ by $2\pi z$, where $z$ is an integer number.
Eq.~\eqref{eq:residual} can also be defined for poloidal mode numbers.
The sign of the mode number determines the direction of propagation in the device frame.
If the helicity (which is determined by the direction of the toroidal magnetic field and the plasma current) is known, the sign of the mode number can be related to the ion or electron diamagnetic drift direction.
By performing the linear fit in each time-frequency point, a time-frequency resolved map of the mode numbers can be generated ~\cite{pokol08experimental}.
Such mode number maps were evaluated to determine the mode numbers of the observed EP-driven modes in section~\ref{sec:results}.

\chapter{Amplitude reconstruction of chirping waves}\label{sec:amp_reconstruction}

For the analysis of EP-driven modes STFT was chosen because it is ideal to examine their time-frequency evolution.
In order to investigate the radial structure, the mode amplitude has to be determined on different SXR LOSs.
STFT provides a good basis for the amplitude reconstruction, because the time-frequency resolution allows to eliminate the major part of the background noise and only the noise component interfering with the mode at the mode frequency causes uncertainty in the reconstruction.
However, to handle the rapidly changing mode frequency and the low signal-to-noise ratio, I developed an advanced reconstruction method which is described in this section.

The derivation of the method is explained in section~\ref{sec:derivation}.
A new approach for interpolating STFT which is necessary for accurate reconstruction is described in section~\ref{sec:interpolation}.
In section~\ref{sec:background}, the effect of background noise on the reconstruction is discussed.
The validation of the method is presented in section~\ref{sec:tests}, where results on synthetic signals are shown.
The linear chirp approximation -- which is finally applied to investigate chirping EP-driven modes -- is investigated in more detail in section~\ref{sec:linear_chirp}.

\section{Derivation of the reconstruction method}\label{sec:derivation}

In order to perform the reconstruction, a signal model is needed.
The fluctuation on the soft X-ray signal caused by the observed mode is modelled as a frequency and amplitude modulated harmonic wave.
The noise of the measured signal is dominated by the soft X-ray light coming from the background plasma.
The noise of the amplifier chain is negligible.
The plasma background noise is modelled as a Gaussian, additive, white noise~$z(t)$.
Since the application of STFT serves as a narrow band filter, the white noise approximation only assumes that the spectral power of the noise is uniform in a narrow frequency range ($\sim10$ kHz).
Thus, this is applicable to model any kind of broadband noise.
The sum of the chirping wave and the additive noise takes the following form:
\begin{equation}\label{eq:signal_model}
  f(t) = a(t)\cos[\phi(t)] + z(t) ,
\end{equation}
where $a(t)$ is the instantaneous amplitude of the wave and the derivative of the phase $\phi(t)$ gives the instantaneous frequency $(\mathrm{d}\phi/\mathrm{d}t = \phi(t)'\equiv \omega(t))$.
During the derivation of the amplitude reconstruction method, the cosine function in eq.~\eqref{eq:signal_model} is substituted by $\exp\{i\phi(t)\}$.
The results can be easily converted considering that
\begin{equation}\label{eq:cosine}
  \cos\big(\phi(t)\big) = \dfrac{\exp\{i \phi(t)\} + \exp\{-i \phi(t)\}}{2} .
\end{equation}
The reconstruction method is derived in the following way.
In eq.~\eqref{eq:signal_model}, $a(t)$ and $\phi(t)$ are expanded in Taylor series around $u$:
\begin{equation}\label{eq:signal_model_taylor_1}
  a_{\mathrm{T}, u}(t) = \sum_{p=0}^{\infty} \frac{a^{(p)}(u)}{p!}(t-u)^p ,
\end{equation}
\begin{equation}\label{eq:signal_model_taylor_2}
  \phi_{\mathrm{T}, u}(t) = \sum_{p=0}^{\infty} \frac{\phi^{(p)}(u)}{p!}(t-u)^p ,
\end{equation}
\begin{equation}\label{eq:signal_model_taylor}
  f_{\mathrm{T}, u}(t) = a_{\mathrm{T}, u}(t)\cos[\phi_{\mathrm{T}, u}(t)] ,
\end{equation}
where $(p)$ denotes the order of the derivation.
Then, eq.~\eqref{eq:signal_model_taylor} is substituted into the definition of STFT (eq.~\eqref{eq:stft}).
At this point noise $z(t)$ is neglected.
The effect of the noise on the results will be discussed in section~\ref{sec:background}.
By evaluating $S f_{\mathrm{T, u}}(u, \xi)$, $a(t)$ can be expressed as a function of the STFT transform and the time derivatives of the amplitude and frequency.
This evaluation was carried out by Oberlin et al.~\cite{oberlin13analysis, oberlin13novel} in the first order of the frequency.
In the present thesis, a more careful treatment of the problem is presented.
In addition to the expansion of the frequency, the time evolution of the amplitude is also taken into account by including its Taylor expansion in eq.~\eqref{eq:signal_model_taylor}.
Furthermore, the effect of higher order terms is also investigated.
Substituting eq.~\eqref{eq:signal_model} into the equation defining STFT (eq.~\eqref{eq:stft}) and separating the zeroth order of the amplitude gives:
\begin{align}\label{eq:stft_taylor_1}
  S f_{\mathrm{T}, u}(u, \xi) &=
  \int\limits_{-\infty}^{+\infty}
  a(u) \exp\{i\phi_{\mathrm{T}, u}(t)\}
  \exp\{-i \xi t\} g(t - u) \dt + \\\nonumber
  &+ \int\limits_{-\infty}^{+\infty}
  \sum_{p=1}^{\infty} \frac{a^{(p)}(u)}{p!}(t-u)^p \exp\{i\phi_{\mathrm{T}, u}(t)\}
  \exp\{-i \xi t\} g(t - u) \dt .
\end{align}
By rearranging this equation, $a(u)$ can be expressed as a function of the STFT transform and the time derivatives of the amplitude and frequency:
\begin{align}\label{eq:stft_taylor_2}
  & a(u) = \\\nonumber
  &= \frac{S f_{\mathrm{T}, u}(u, \xi) - \displaystyle\int\limits_{-\infty}^{+\infty}
  \sum_{p=1}^{\infty} \frac{a^{(p)}(u)}{p!}(t-u)^p \exp\{i\phi_{\mathrm{T}, u}(t)\}
  \exp\{-i \xi t\} g(t - u) \dt}
  {\displaystyle\int\limits_{-\infty}^{+\infty} \exp\{i\phi_{\mathrm{T}, u}(t)\}  \exp\{-i \xi t\} g(t - u) \dt} .
\end{align}
Eq.~\eqref{eq:stft_taylor_2} gives an exact formula to evaluate the instantaneous amplitude if the background noise power is zero.
However, this formula requires the knowledge of the instantaneous frequency and its derivatives ($\phi^{(p)}(u)$) and the derivatives of the amplitude ($a^{(p)}(u)$, $p\geq1$).
Therefore, eq.~\eqref{eq:stft_taylor_2} is a recursive definition of $a(u)$, because the knowledge of the amplitude is required to calculate its derivatives.
An estimation for the frequency and the amplitude can be given considering eq.~\eqref{eq:stft_taylor_1} in the zeroth order for the amplitude and in the first order for the phase:
\begin{equation}\label{eq:stft_taylor_zeroth_1}
  S f_{\mathrm{T}, u}(u, \xi) =
  \int\limits_{-\infty}^{+\infty}
  a(u) \exp\{i\phi(u) + i\phi'(u)(t-u)\}
  \exp\{-i \xi t\} g(t - u) \dt .
\end{equation}
This formula gives exact instantaneous amplitude in the case of constant amplitude and constant frequency waves.
Considering the formula of the $g(t)$ Gaussian window function from eq.~\eqref{eq:gabor_atom} the integral in eq.~\eqref{eq:stft_taylor_zeroth_1} can be analytically evaluated:
\begin{equation}\label{eq:stft_taylor_zeroth_2}
  S f_{\mathrm{T}, u}(u, \xi) = 
  a(u) \sqrt{2\sigma_t\sqrt{\pi}} \exp\{ i\phi(u) -i\xi u \} \exp\left\{ -\frac{\sigma_t^2}{2}(\phi'(u) - \xi)^2 \right\} .
\end{equation}
It is clear from eq.~\eqref{eq:stft_taylor_zeroth_2} that $|S f_{\mathrm{T}, u}(u, \xi)|$ has its maximum where ${\phi'(u)=\xi}$.
Therefore, the frequency of the mode can be estimated by searching for the maxima on the spectrogram.
It follows from eq.~\eqref{eq:stft_taylor_zeroth_2} that the mode amplitude can be evaluated from the maximum value of $|S f_{\mathrm{T}, u}(u, \xi)|$:
\begin{equation}\label{eq:stft_taylor_zeroth_3}
  a(u) = \frac{|S f_{\mathrm{T}, u}(u, \xi = \phi'(u))|}{\sqrt{2\sigma_t\sqrt{\pi}}} .
\end{equation}
By using this approximation of the amplitude and frequency, the derivatives in eq.~\eqref{eq:stft_taylor_2} can be estimated and $a(u)$ can be evaluated numerically.

In principle, the determination of $p$ in eq.~\eqref{eq:stft_taylor_2} above which the higher order terms can be neglected, depends on the behaviour of the chirping wave.
The Gaussian window (eq.~\eqref{eq:gabor_atom}) applied for STFT has a well-localized energy in time which time scale is characterized by the $\sigma_t$ standard deviation.
Eq.~\eqref{eq:stft_taylor_2} can give a good approximation with a limited $p=p_{\mathrm{max}}$ if the remaining terms $p<p_{\mathrm{max}}$ accurately describe the chirping wave on the time scale of the window width $\sigma_t$.
For example, if a linear chirp model (first order in amplitude and second order in phase) was a good approximation of the observed wave on the time scale of $\sigma_t$, then the higher order terms would give negligible contribution to the result.

In summary, the amplitude reconstruction method -- if higher order terms are taken into consideration -- comprises the following steps:
First, the instantaneous frequency of the mode is estimated by using a maximum-searching algorithm.
The time evolution of the maxima of the mode frequency is traced, which is called the ridge of the STFT transform.
Second, by using the zeroth order approximation (eq.~\eqref{eq:stft_taylor_zeroth_3}) the instantaneous amplitude is estimated from the STFT values of the ridge.
Then, the derivatives of the amplitude $(a^{(p)}(u))$ and frequency ($\phi^{(p)}(u)$) are evaluated from the zeroth order approximation.
Finally, the instantaneous amplitude is approximated by calculating eq.~\eqref{eq:stft_taylor_2}.
In order to avoid the effect of quantization error in the numerical derivatives of the amplitude and frequency, an interpolation of STFT is required.

\section{Interpolation of STFT}\label{sec:interpolation}

The interpolation method that I used in this thesis could have a broader applicability than a simple interpolation algorithm.
With this technique the reconstruction of the continuous STFT of a signal at any time-frequency point is possible.
This method is based on a theorem introduced in the book of Mallat~\cite{mallat08wavelet}.
Let $\Psi$ be an element of square integrable functions ($\Psi\in L^2(\mathbb{R}^2)$).
There exists $f\in L^2(\mathbb{R})$ such that $\Psi(u,\xi)=Sf(u,\xi)$, if and only if,
\begin{equation}\label{eq:interpol}
  \Psi(u_0, \xi_0) = \dfrac{1}{2 \pi} \int\limits_{-\infty}^{\infty} \int\limits_{-\infty}^{\infty}
  \Psi(u, \xi) K(u_0, u, \xi_0, \xi) \mathrm{d}u \mathrm{d}\xi ,
\end{equation}
where
\begin{equation}
  K(u_0, u, \xi_0, \xi) = \langle g_{u, \xi}, g_{u_0, \xi_0} \rangle
\end{equation}
is the reproducing kernel.
This theorem shows that not any two dimensional distribution can be interpreted as the STFT of some $f$ function.
If the discretization of the STFT is handled carefully as it is described in section~\ref{sec:stft}, then the integral in eq.~\eqref{eq:interpol} can be well approximated by the discrete STFT.
In this way the continuous transform can be reconstructed at any given time-frequency point.

\begin{figure}[htb!]\centering
  \includegraphics[width = 0.75\textwidth]{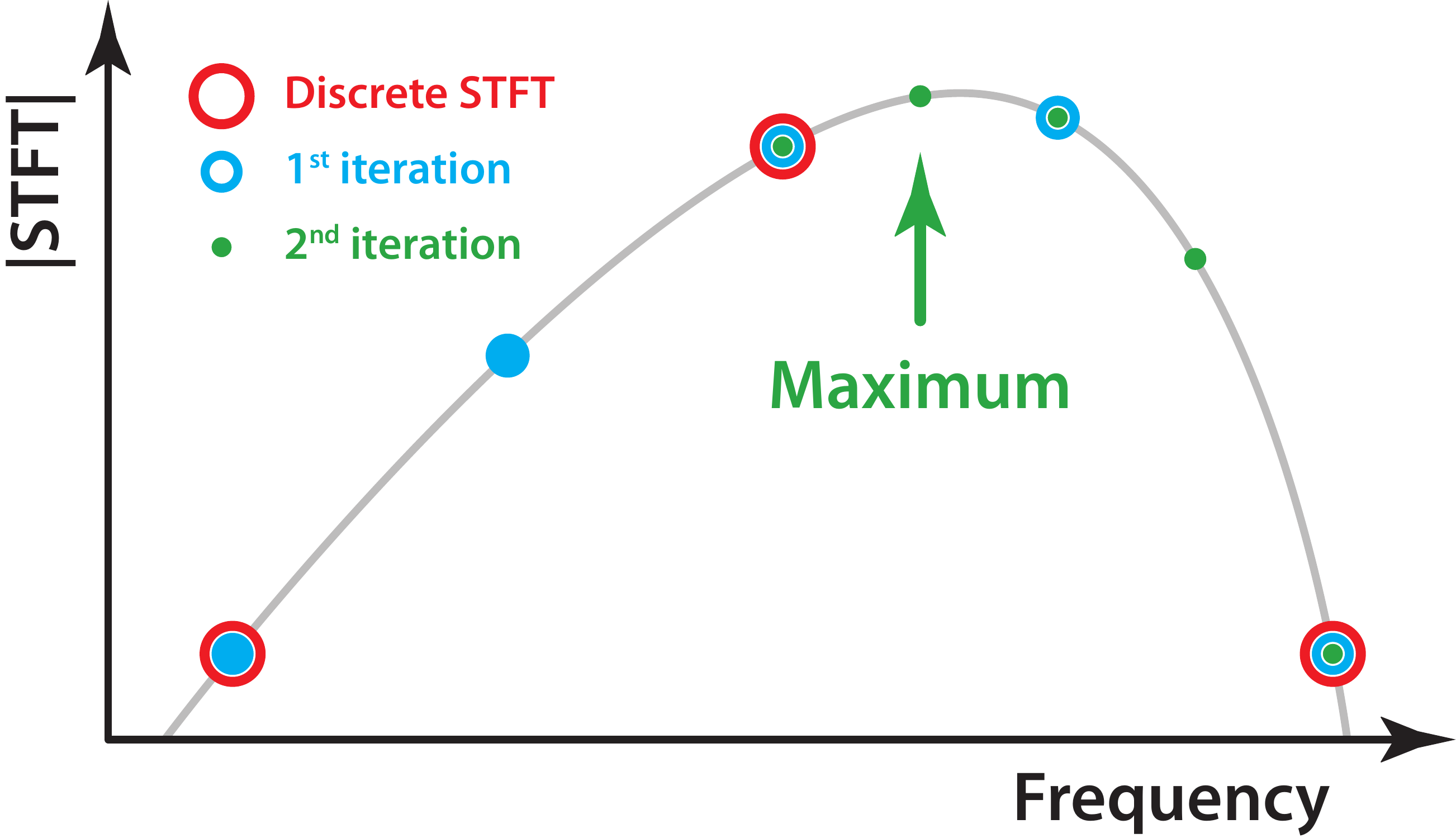}
  \caption{Illustration of the interpolation method which was used to accurately evaluate the STFT ridge.}
  \label{fig:interpol_demo}
\end{figure}

I searched for the local maxima of the absolute value of the STFT at each time-point by interpolating the transform in frequency.
The method is illustrated in figure~\ref{fig:interpol_demo}.
The red circles show the values of the discrete STFT and the grey solid line presents the continuous transform.
The transform is interpolated in the frequency points halfway between the red points.
This gives the cyan points which are the results of the first iteration of the interpolation.
Then, the highest value from the cyan points is selected.
After that the transform is interpolated again halfway between the neighbouring cyan points.
This gives the green points which determine a more accurate estimation for the STFT ridge.
In practice, this iteration was repeated $10-15$ times to avoid quantization error in the numerical time derivatives of the amplitude and frequency.

\section{Background noise level}\label{sec:background}

\begin{figure}[htb!]\centering
  \includegraphics[width = 1.0\textwidth]{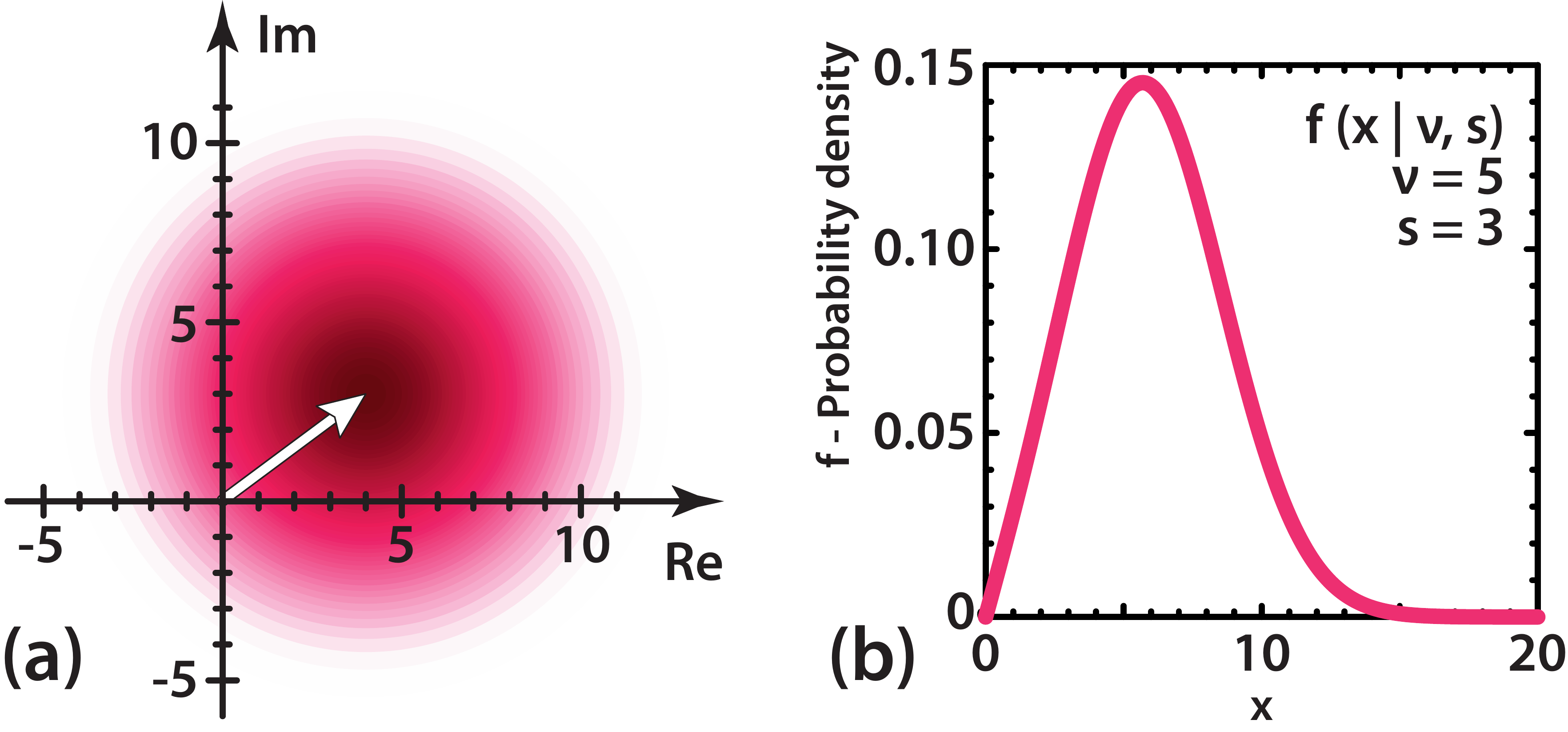}
  \caption{\textbf{(a)} This figure illustrates the sum of the STFT transform of a deterministic wave (white vector with length of $\nu=5$) and a Gaussian, additive, white noise (pink spot with standard deviation $\sigma=3$) at a given time-frequency point $(u, \xi)$. \textbf{(b)} The probability density of the Rice distribution with parameters $\nu=5$ and $s=3$.}
  \label{fig:noise_demo}
\end{figure}

The derivation presented in section~\ref{sec:derivation} is valid if there is no additional noise in the signal.
Thus, the effect of the noise has to be investigated.
As it was mentioned before, my signal model (eq.~\eqref{eq:signal_model}) contains a Gaussian, additive, white noise:
\begin{equation}\label{eq:signal_model_2}
  f(t) = a(t)\cos[\phi(t)] + z(t) = f_{\mathrm{chirp}}(t) + z(t) .
\end{equation}
Since STFT is a linear transform:
\begin{equation}\label{eq:linearity}
  S \big( f_{\mathrm{chirp}}(t) + z(t) \big) = S f{_\mathrm{chirp}}(t) + S z(t) .
\end{equation}
To illustrate the problem, the transform of the terms in eq.~\eqref{eq:linearity} are calculated separately, then the behaviour of their sum is investigated on the complex plane.
At a given time-frequency point $(u, \xi)$ the value of the transform of $f{_\mathrm{chirp}}(t)$ gives a complex number whose amplitude and phase are deterministic.
This complex number is plotted as a white vector on the complex plane in figure~\ref{fig:noise_demo}a.
The STFT of the Gaussian, additive, white noise $z(t)$ at a given time-frequency point $(u, \xi)$ is a complex number whose amplitude is proportional to the noise power, and its phase is an uniformly distributed random number.
The distribution of this vector is illustrated with the pink spot on the complex plane in figure~\ref{fig:noise_demo}a centred around the deterministic component.

Since the STFT of the Gaussian white noise has a 2D Gaussian distribution on the complex plane, the magnitude of the sum of the two vectors has a Rice distribution~\cite{dekker14data}.
The Rice or Rician distribution is the probability distribution of the magnitude of a circular bivariate normal random variable with potentially non-zero mean~\cite{rice45mathematical}.
Its probability density function is the following:
\begin{equation}\label{eq:rice}
  f(x | \nu, s) = \frac{x}{s^2} \exp\bigg\{ \frac{-(x^2 + \nu^2)}{2s_2} \bigg\} I_0\bigg(\frac{x\nu}{s^2}\bigg) ,
\end{equation}
where $\nu$ is the distance of the centre of the 2D Gaussian distribution from the origin, $s$ is the standard deviation of the 2D Gaussian distribution and $I_0(z)$ is the modified Bessel function of the first kind with order zero.
The probability density of the Rice distribution with parameters $\nu=5$ and $s=3$ is shown in figure~\ref{fig:noise_demo}b.
The mean of a random variable $x$ with Rice distribution is
\begin{equation}\label{eq:rice_mean}
  \mathrm{E}[x] = s\sqrt{\pi/2}\ L_{\rfrac{1}{2}}(-\nu^2/2s^2)
\end{equation}
and its variance is
\begin{equation}\label{eq:rice_var}
  \mathrm{Var}[x] = 2s^2 + \nu^2 - \frac{\pi s^2}{2} L^2_{\rfrac{1}{2}}(-\nu^2/2s^2) ,
\end{equation}
where $L_q(x)$ denotes a Laguerre polynomial.

The background noise level, i.e. the $s$ of the 2D Gaussian distribution is determined in the following way.
A time-frequency interval is defined on the spectrogram of the given signal where no coherent mode appears.
Since it is assumed that only the Gaussian white noise gives significant contribution to the values over the selected time-frequency interval, the real or imaginary part of these values has a Gaussian distribution with standard deviation $s$, where $s$ is the standard deviation of the 2D Gaussian distribution of the noise.
Thus, the background noise level is estimated by calculating the standard deviation of the real part of the STFT values over the selected time-frequency interval.

As $\nu$ becomes large or $s$ becomes small the mean of the Rice distribution becomes $\nu$ and the variance becomes $s^2$~\cite{kushal91estimation}.
In the investigated real cases, the ratio of $\nu$ and $s$ is around~$10$ where the oscillation amplitude is maximal.
Therefore, in the present thesis, I use the $\nu \gg s$ limit which gives a good approximation to estimate the uncertainty of the reconstructed amplitude.
In figures showing the reconstructed amplitude in section~\ref{sec:results}, the error bars were derived from the standard deviation $s$ of the background noise by taking into account the error propagation in the reconstruction formula.

\section{Numerical tests}\label{sec:tests}

\begin{figure}[htb!]\centering
  \includegraphics[width = 1.0\textwidth]{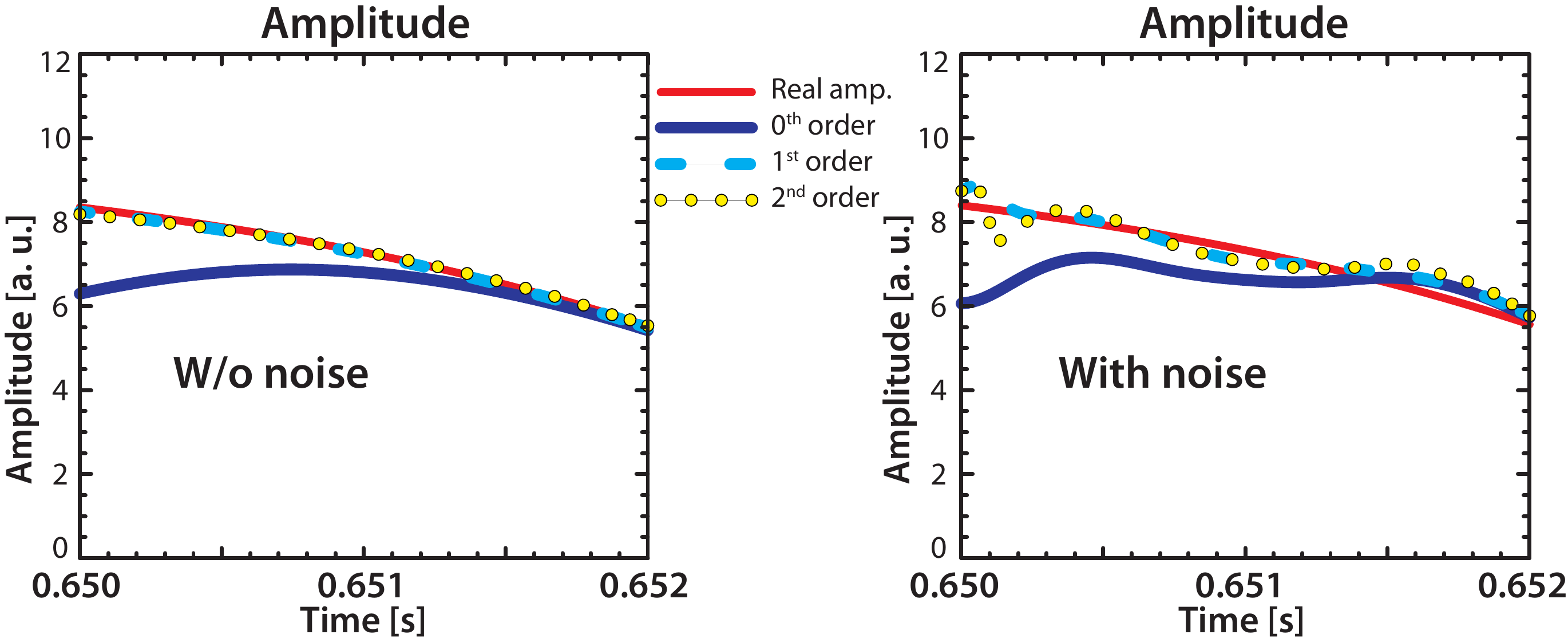}
  \caption{\textbf{(a)} The result of amplitude reconstruction taking into account higher order terms for a synthetic signal without noise. \textbf{(b)} Amplitude reconstruction for synthetic signal with additive Gaussian white noise.}
  \label{fig:test_synth}
\end{figure}

I performed numerical tests to investigate the effect of the higher order terms and the background noise on the amplitude reconstruction.
These tests were carried out on synthetic signals which had similar frequency and amplitude time evolution as the measured signals.
The amplitude and frequency of the generated signals presented in this thesis were the following:
\begin{equation}
  \omega(t) = 2\pi\cdot10^4\big(\exp\{ -10^{-3}t \} + 7\big)\ \mathrm{[Hz]},
\end{equation}
\begin{equation}
  a(t) = -\exp\{ 5\cdot10^2t \} + 10 .
\end{equation}
First, the tests were carried out without adding noise to the signal to investigate the effect of higher order terms in the amplitude reconstruction.
In figure~\ref{fig:test_synth}a the pre-defined amplitude of the synthetic chirp is shown with red solid line.
The other lines show the reconstruction taking into account increasing number of higher order terms.
Maximum order corresponds to the highest derivative of the amplitude and frequency taken into account.
Figure~\ref{fig:test_synth}a shows that the first order approximation gives a better estimation of the pre-defined amplitude evolution.
However, the second order terms do not give significant contribution to the results.

In figure~\ref{fig:test_synth}b the effect of additive noise on the reconstruction is examined.
In this case an additive Gaussian white noise was added to the synthetic signal.
The signal-to-noise ratio had similar magnitude as in the investigated real measurements.
For the synthetic signal, this means that the amplitude of the wave was around $8$ and the standard deviation of the additive Gaussian noise was set to be $15$.
These tests had already shown that the STFT based method effectively filters out the noise and can deal with this very poor signal-to-noise ratio.

The noise causes higher uncertainty in the zeroth order estimation of the amplitude and frequency.
This uncertainty leads to high errors when the derivatives are calculated.
As it is visible in figure~\ref{fig:test_synth}b, the first order approximation gives a better estimation for the pre-defined amplitude than the zeroth order approximation.
However, the effect of the noise is more significant in the case of the second order approximation due to the higher order derivatives.

The results of tests on synthetic signals suggest that the first order approximation estimates well the amplitude of the synthetic chirp.
This is a trade-off between the magnitude of the correction given by the higher order terms and the uncertainty arising from the derivatives of the noisy zeroth order approximation of the amplitude and the frequency.
Since the tests have shown that the first order approximation is sufficient for my purposes, this is investigated in more detail.

\section{Linear chirp approximation}\label{sec:linear_chirp}

Eq.~\eqref{eq:stft_taylor_1} in the first order for the amplitude and second order for the phase gives
\begin{align}\label{eq:stft_linear_1}
  & S f_{\mathrm{T}, u}(u, \xi) = \\\nonumber
  & = \int\limits_{-\infty}^{+\infty}
  \big(a(u) + a'(u)(t-u)  \big) \exp\Big\{i \Big( \phi(u) + \phi'(u)(t-u) + \frac{\phi''(u)}{2}(t-u)^2 \Big) \Big\} \\\nonumber
  & \exp\{-i \xi t\} \frac{1}{\sqrt{\sqrt{\pi}\sigma_t}}\exp\left\{-\frac{(t-u)^2}{2\sigma_t^2}\right\} \dt .
\end{align}
Even in case of a linear chirp, the absolute value of the transform has its maximum where $\phi'(u) = \xi$~\cite{oberlin13analysis}.
Evaluating eq.~\eqref{eq:stft_linear_1} and taking into account that $\phi'(u) = \xi$ yields
\begin{align}\label{eq:stft_linear_2}
  S f_{\mathrm{T}, u}&(u, \xi) =
  \frac{1}{\sqrt{\sqrt{\pi}\sigma_t}} \exp\Big\{i\Big(\phi(u) - \xi u\Big)\Big\} \\\nonumber
  & \Bigg[ a(u) \int\limits_{-\infty}^{+\infty}
  \exp\bigg\{ \bigg( \frac{i\phi''(u)}{2}-\frac{1}{2\sigma_t^2} \bigg) (t-u)^2 \bigg\} \dt + \\\nonumber
  + & a'(u) \int\limits_{-\infty}^{+\infty}
  (t-u) \exp\bigg\{ \bigg( \frac{i\phi''(u)}{2}-\frac{1}{2\sigma_t^2} \bigg) (t-u)^2 \bigg\} \dt \Bigg] .
\end{align}
By performing variable substitution in the integral of the second term for $t-u$, it is clearly visible that this integral contains the product of an odd and an even function.
Thus, the second term equals to zero which shows that the $a'(u)$ does not give contribution to the result.
The first integral can be evaluated analytically, thus $a(t)$ can be expressed as
\begin{equation}\label{eq:stft_linear_3}
  a(u) = \sqrt{\frac{1 - i\phi''(u)\sigma_t^2}{2\sigma_t\sqrt{\pi}}} \exp\{ -i(\phi(u) - \xi u) \} S f_{\mathrm{T}, u}(u, \xi) .
\end{equation}
Taking into consideration that $a(u)$ is a real number, its value is given by the absolute value of eq.~\eqref{eq:stft_linear_3}:
\begin{equation}\label{eq:stft_linear_4}
  \boxed{a(u) = \frac{| S f_{\mathrm{T}, u}(u, \xi) |}{\sqrt{2\sigma_t\sqrt{\pi}}} \sqrt[4]{1 + \phi''(u)^2\sigma_t^4} .}
\end{equation}
With this formula the amplitude of the wave can be reconstructed when the linear chirp is a good approximation on the time scale of $\sigma_t$.
This formula differs from the zeroth order approximation (eq.~\eqref{eq:stft_taylor_zeroth_3}) in the following correction factor: $\sqrt[4]{1 + \phi''(u)^2\sigma_t^4}$.
This correction depends on the derivative of the frequency and the width of the applied Gaussian window which determines the time-frequency resolution of STFT.
It is clear that the correction factor is lower if the window width is smaller, i.e when the time resolution of the transform is better.
However, the window width cannot be arbitrary small, because the better the time-resolution the worse the frequency resolution.
Therefore, a trade-off between the time- and frequency resolution is needed.
In a previous work I used the zeroth order approximation, but the time-frequency resolution was set to minimize the effect of the neglected correction~\cite{horvath14changes}.
In this thesis I used the first order approximation, because it allows to apply a wider window function which provides better frequency resolution, thus more accurate reconstruction of the STFT ridge.

\begin{figure}[htb!]\centering
  \includegraphics[width = 1.0\textwidth]{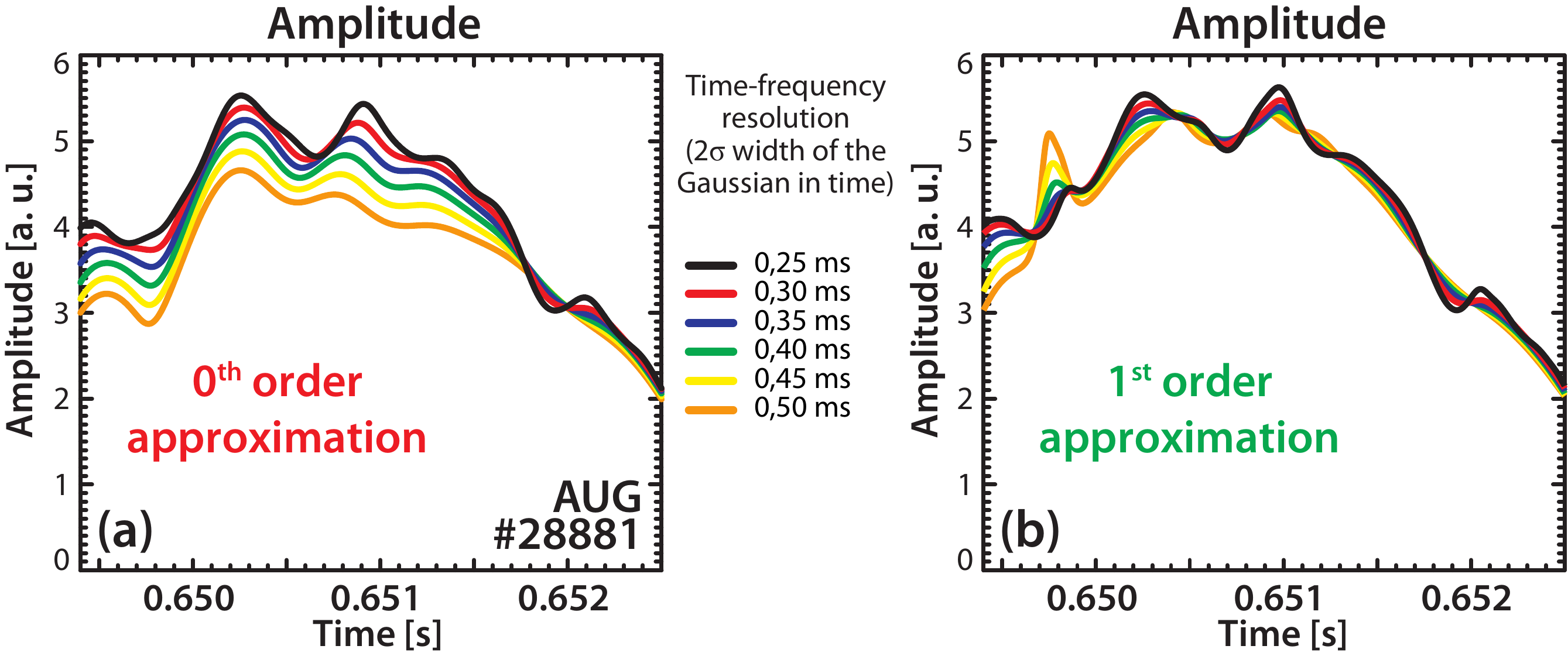}
  \caption{\textbf{(a)} The result of amplitude reconstruction by using the zeroth order approximation (eq.~\eqref{eq:stft_taylor_zeroth_3}) on a real signal. \textbf{(b)} The result of amplitude reconstruction by using linear chirp approximation (eq.~\eqref{eq:stft_linear_4}) on a real signal.}
  \label{fig:test_real}
\end{figure}

Figure~\ref{fig:test_real} shows the effect of the time-frequency resolution on the reconstruction.
To confirm that the linear chirp approximation gives the best estimation for the chirping modes investigated in real signals, in this case I reconstructed the amplitude of a real oscillation (BAE) with different time-frequency resolution.
In figure~\ref{fig:test_real}a, the zeroth order approximation (eq.~\eqref{eq:stft_taylor_zeroth_3}) is used.
As it is visible, the reconstructed amplitude depends on the time-frequency resolution.
Figure~\ref{fig:test_real}b shows the case of reconstruction with the linear chirp approximation (eq.~\eqref{eq:stft_linear_4}).
In this case the time-frequency resolution has a much weaker effect on the reconstruction.

As it was discussed in this section, with an accurate signal model, the amplitude of rapidly chirping waves can be reconstructed.
The effect of additive noise is handled separately by taking advantage of the linearity of the STFT.
Numerical tests had shown that the linear chirp approximation gives a good approximation for the amplitude of chirping waves which have similar parameters as the ones observed in the measurement.
Therefore, in the following section, where the experimental observations are discussed, the first order approximation (eq.~\eqref{eq:stft_linear_4}) is applied.

\chapter{Experimental observation of EP-driven modes}\label{sec:results}

The importance of EP-driven modes regarding future burning plasma experiments was discussed in section~\ref{sec:introduction}.
Present section shows a detailed analysis to experimentally characterize the behaviour of EP-driven plasma modes.
The diagnostic tools presented in section~\ref{sec:setup} and data processing methods developed in section~\ref{sec:amp_reconstruction} allow to examine the rapid changes in the radial structure of bursting EP-driven modes during the non-linear chirping phase.

The experiments in question were carried out on ASDEX Upgrade (AUG).
The scenario in which BAEs and EGAMs were observed was characterized by low density ($\sim2\cdot10^{19}\ \mathrm{m}{}^{-3}$) in order to maximize the fast particle pressure fraction.
The modes were observed in the current ramp-up phase of the discharge when off-axis NBI with 93 keV beam energy was applied.
The ramp-up phase in a typical AUG discharge lasts about 1 sec.
A series of discharges with strong Alfvén activity were investigated.
The list of analysed discharges with the approximate time and frequency intervals where BAEs and EGAMs were observed is presented in table~\ref{tab:shots}.
\begin{table}[htb!]
\footnotesize
  \begin{center}
    \begin{tabular}{|c||c|c||c|c|}
\hline
    &\multicolumn{2}{c||}{BAEs}	&\multicolumn{2}{c|}{EGAMs}	\\
	&Time [s]	&Frequency [kHz]	&Time [s]	&Frequency [kHz]	\\
\hline\hline
\#25506	&0.45 - 0.60	&50 - 60	&$\times$	&$\times$\\
\#28881	&0.64 - 0.72	&70 - 80	&0.85 - 1.00	&60 - 80\\
\#28884	&0.62 - 0.72	&70 - 85	&$\times$	&$\times$\\
\#28885	&$\times$	&$\times$	&0.75 - 0.95	&70 - 85\\
\#30383	&$\times$	&$\times$	&0.25 - 0.60	&30 - 60\\
\#30945	&$\times$	&$\times$	&$\times$	&$\times$\\
\#30946	&0.45 - 0.60	&40 - 70	&0.60 - 0.70	&40 - 60\\
\#30950	&0.45 - 0.55	&50 - 60	&0.70 - 0.80	&60 - 80\\
\#30951	&0.55 - 0.70	&70 - 90	&$\times$	&$\times$\\
\#30952	&0.40 - 0.45	&60 - 80	&0.50 - 0.70	&50 - 70\\
\#30953	&$\times$	&$\times$	&0.55 - 0.75	&40 - 60\\
\#31213	&$\times$	&$\times$	&0.65 - 1.00	&40 - 60\\
\#31214	&$\times$	&$\times$	&0.45 - 0.80	&60 - 80\\
\#31215	&$\times$	&$\times$	&0.45 - 0.75	&40 - 60\\
\#31216	&$\times$	&$\times$	&0.50 - 0.60	&50 - 70\\
\#31233	&$\times$	&$\times$	&0.80 - 1.00	&60 - 80\\
\#31234	&$\times$	&$\times$	&0.65 - 0.80	&40 - 50\\
\hline
    \end{tabular}
    \caption{\label{tab:shots} Investigated shots with strong Alfvén activity. The approximate time and frequency intervals where BAEs and EGAMs were observed are presented.}
  \end{center}
\end{table}
The mode type was identified mainly by considering the toroidal mode number.
EGAMs have a toroidal mode number $n = 0$ and BAEs have a nonzero toroidal mode number.

The main goal of this thesis is to investigate the changes in the radial structure of the mode in the non-linear chirping phase.
The only fluctuation measurement which has spatial resolution and these modes were visible in its signal was the SXR diagnostic.
Since the SXR diagnostic is a line-integrated measurement, it is not straightforward to reconstruct the radial structure of the mode, but many line-of-sights (LOSs) are available which makes it possible to qualitatively investigate the time evolution of the radial structure.

The identification of the modes was carried out by using the magnetic pick-up coils.
Magnetic spectrograms are excellent to follow the time evolution of the modes and determine mode numbers.
Note that in principle EGAMs should only be detectable with density fluctuation measurements because it is and electrostatic mode, however, due to sideband coupling it is clearly visible on magnetic fluctuation measurements as well~\cite{lauber13super}.

In general, the signal-to-noise ratio of the SXR measurements was poor.
In many cases, modes which were clearly visible on the magnetic spectrogram were only visible on one or two LOSs of SXR.
The strategy was to find cases where the mode is observable on at least 3 adjacent LOSs of a particular SXR camera.
LOSs of SXR cameras F, G, H, I and J (see section~\ref{sec:sxr}) were investigated in the discharges listed in table~\ref{tab:shots} by calculating the spectrogram of their signal in the time-frequency range where the mode was visible on the magnetic spectrogram.
In total, about $1000$ SXR spectrograms were analysed in this process.
Finally, 3 cases for BAEs and 5 cases for EGAMs were found where the oscillation amplitude on the SXR signals was sufficiently high to identify them visually on the SXR spectrograms.

\section{Beta-induced Alfvén eigenmodes}

\begin{figure}[htb!]\centering
  \includegraphics[width = 0.75\textwidth]{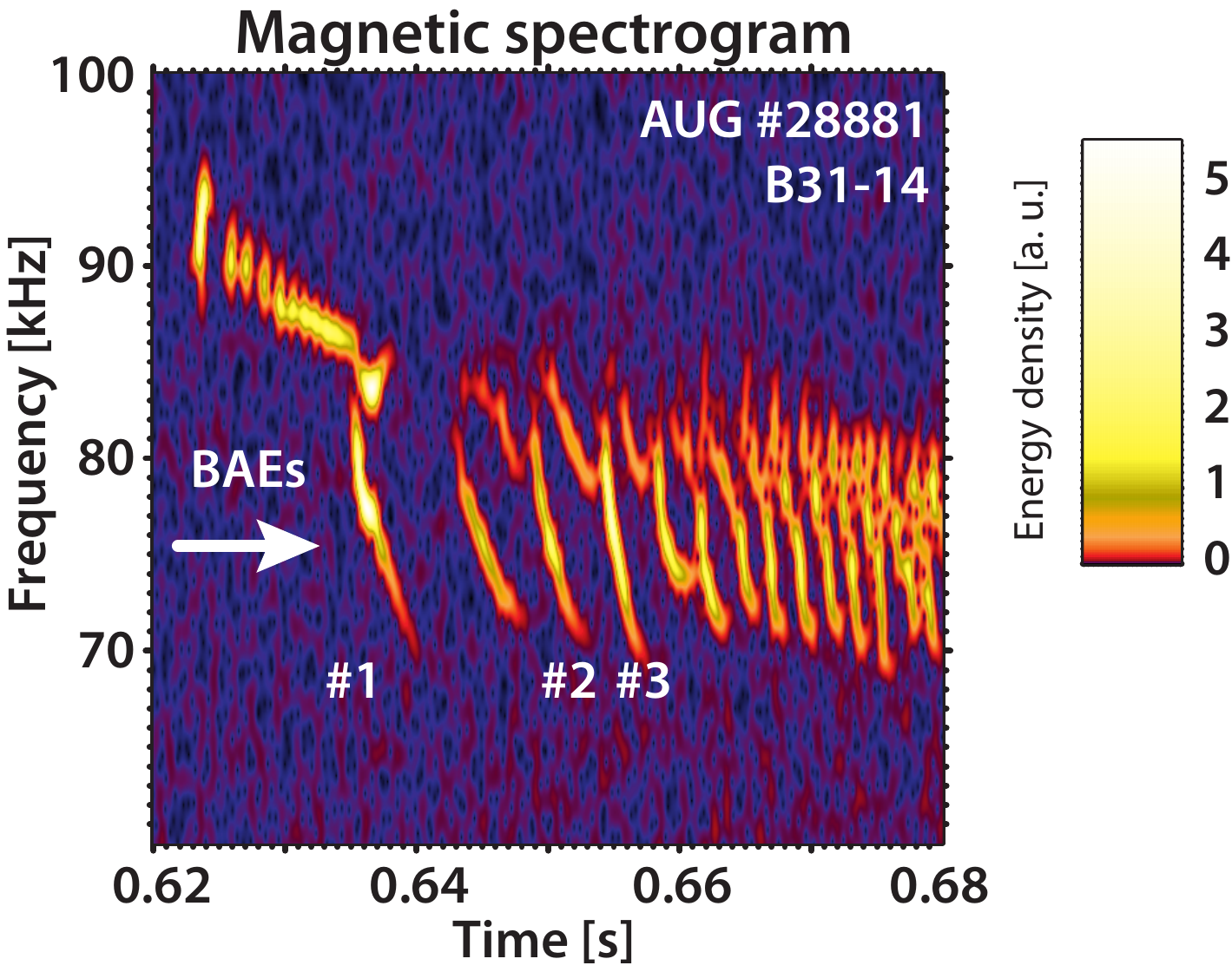}
  \caption{Chirping BAEs with decreasing frequency in the range of $70-85$ kHz are visible on the magnetic spectrogram. Numbers denote the chirps which were investigated in detail. Modes around $90$ kHz were not identified.}
  \label{fig:bae_magnetic}
\end{figure}

First, the results from the observation of beta-induced Alfvén eigenmodes (BAEs) are presented.
As it is shown in table~\ref{tab:shots}, BAEs were observed in seven discharges.
However, the signal-to-noise ratio of the SXR signals was only appropriate for further analysis in shot \#28881.
In this shot on-axis NBI heating was applied early (from $t = 0.35$ s), then off-axis NBI was injected at $t = 0.6$ s.
Shortly after, bursting BAEs appeared at around 80 kHz as it is shown on the magnetic spectrogram on figure~\ref{fig:bae_magnetic}.
The modes are observed to chirp down in frequency by as much as 10 kHz in 2-3 ms.

\subsection{Mode number analysis of BAEs}

\begin{figure}[htb!]\centering
  \includegraphics[width = 1.0\textwidth]{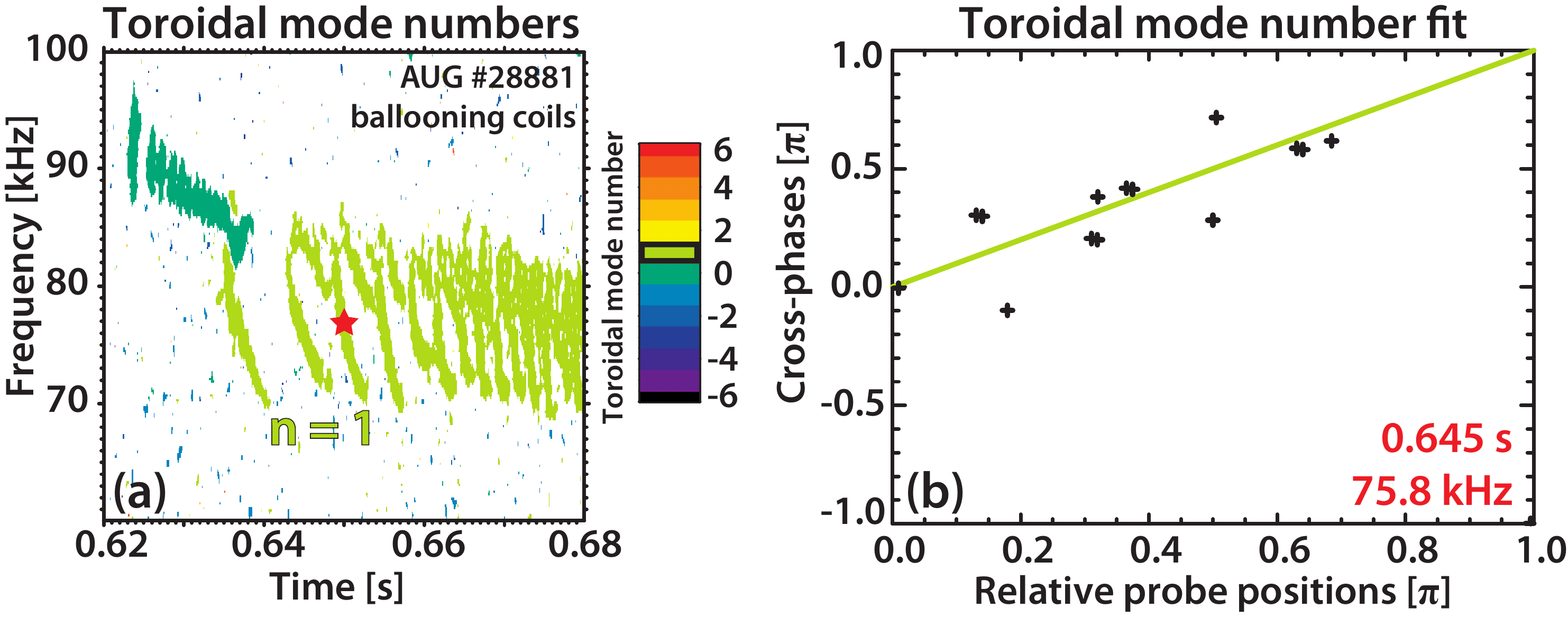}
  \caption{Toroidal mode number calculation for BAEs. \textbf{(a)} The result of time-frequency resolved mode number calculation. The mode numbers are plotted only in time-frequency points where the residual of the fit is lower than 5~\% of the maximum. \textbf{(b)} The relative phases between all pairs of signals are plotted as a function of relative probe position in the time-frequency point marked with a red star in figure~\ref{fig:bae_toroidal}a. The best fitting line with $n = 1$ slope is plotted with light green solid line.}
  \label{fig:bae_toroidal}
\end{figure}

An important step in mode identification is determining the toroidal and poloidal mode numbers.
The toroidal mode number was determined from the signals with correction~\cite{horvath15reducing} of the ballooning coils as it was described in section~\ref{sec:modenumber}.
The result of the toroidal mode number evaluation is presented in figure~\ref{fig:bae_toroidal}.
In figure~\ref{fig:bae_toroidal}a, the time-frequency resolved mode number calculation is shown where it is visible that BAEs have $n=1$ toroidal mode number.
The mode numbers are plotted only in time-frequency points where the residual of the fit is lower than the 5~\% of the maximum.
This filter ensures that only well-fitting mode numbers are taken into account.
In figure~\ref{fig:bae_toroidal}b, the relative phases between all pairs of signals are plotted as a function of relative probe position in the time-frequency point marked with a red star in figure~\ref{fig:bae_toroidal}a.
The best fitting line with $n = 1$ slope is plotted with black solid line.
\begin{figure}[htb!]\centering
  \includegraphics[width = 1.0\textwidth]{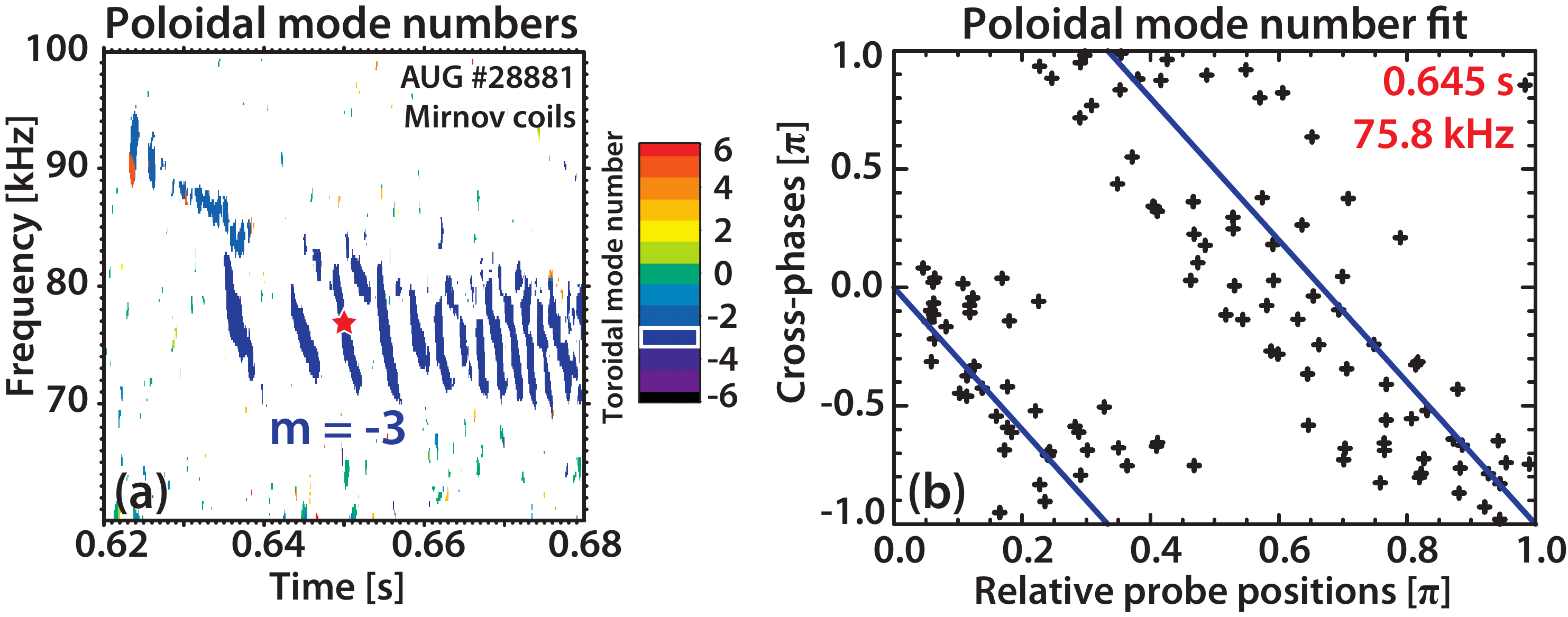}
  \caption{Poloidal mode number calculation for BAEs. \textbf{(a)} The result of time-frequency resolved mode number calculation. The mode numbers are plotted only in time-frequency points where the value of the minimum coherence is higher than~0.3. \textbf{(b)} The relative phases between all pairs of signals are plotted as a function of relative probe position in the time-frequency point marked with a red star in figure~\ref{fig:bae_poloidal}a. The best fitting line with $n = -3$ slope is plotted with blue solid lines.}
  \label{fig:bae_poloidal}
\end{figure}

In order to determine the poloidal mode number, the poloidal array of Mirnov coils was used.
Despite the difficulties concerning the use of Mirnov-coils (see section~\ref{sec:magnetic}), the results presented in figure~\ref{fig:bae_poloidal} clearly show that the poloidal mode number of the investigated BAEs is $m=-3$.
The negative sign means that the mode propagates in the ion diamagnetic drift direction.
Since the Mirnov coils are not calibrated, in figure~\ref{fig:bae_poloidal}a a different filter is applied than in the case of toroidal mode numbers: the mode numbers are plotted only in time-frequency points where the value of the minimum coherence~\cite{pokol10wavelet} is higher than~0.3 for number of averages of 3.
Figure~\ref{fig:bae_poloidal}b shows the relative phases between all pairs of signals as a function of relative probe position.

\subsection{Changes in the radial structure of BAEs}

\begin{figure}[htb!]\centering
  \includegraphics[width = 1.0\textwidth]{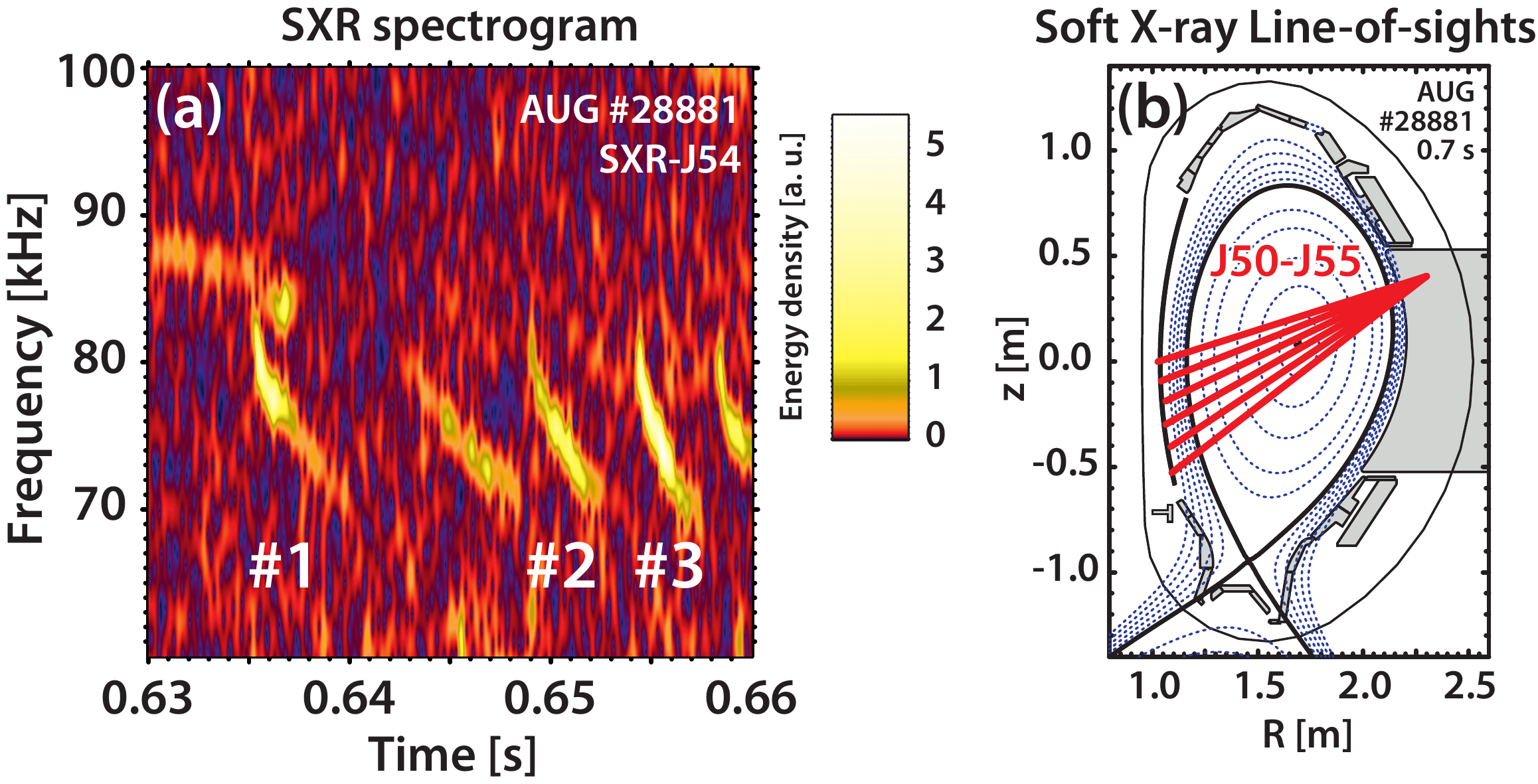}
  \caption{\textbf{(a)}~Spectrogram calculated from the signal of SXR LOS J-54. The downchirping BAEs are also visible here similarly to the magnetic spectrogram (figure~\ref{fig:bae_magnetic}, note the different time axis.). \textbf{(b)}~Six LOSs of SXR camera J (red lines) where the signal-to-noise ratio was sufficiently high for further analysis.}
  \label{fig:bae_sxr_spectrogram}
\end{figure}

The radial structure analysis of BAEs was carried out by using the LOSs of SXR camera J.
These LOSs are shown in figure~\ref{fig:bae_sxr_spectrogram}b with red lines.
Based on the signal-to-noise ratio of the SXR signals three chirps were selected for further analysis.
The soft X-ray spectrogram of channel J54 is shown in figure~\ref{fig:bae_sxr_spectrogram}a.
The instantaneous amplitude of the mode on each LOS is calculated by using the first order approximation defined in eq.~\eqref{eq:stft_linear_4}.
Since eq.~\eqref{eq:stft_linear_4} contains the derivative of the instantaneous frequency, the precise determination of the instantaneous amplitude requires an accurate frequency ridge.
Therefore, the frequency ridge was evaluated from the magnetic spectrogram, because the magnetic signals have a higher signal-to-noise ratio.
The ridge was traced by a local maximum searching algorithm~\cite{horvath14changes, papp11low}, which follows the mode evolution until the mode amplitude falls below the background noise level.

\begin{figure}[htb!]\centering
  \includegraphics[width = 1.0\textwidth]{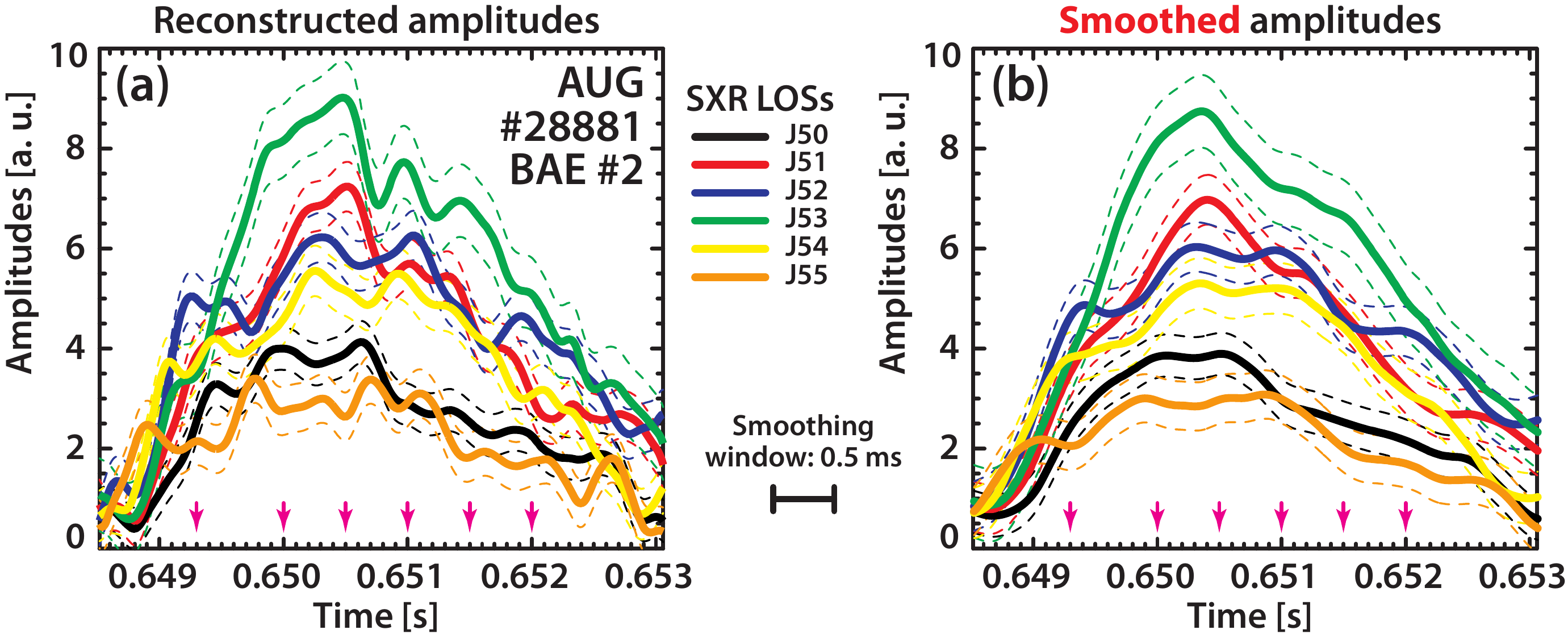}
  \caption{The time evolution of the oscillation amplitude of BAE chirp \#2 (see figure~\ref{fig:bae_magnetic}) on the different SXR LOSs. \textbf{(a)}~Reconstructed amplitudes by using linear chirp approximation (eq.~\eqref{eq:stft_linear_4}). \textbf{(b)}~Amplitudes smoothed by a moving average with boxcar kernel of $0.5$~ms width.}
  \label{fig:bae_amps_maxorder1}
\end{figure}
\begin{figure}[htb!]\centering
  \includegraphics[width = 1.0\textwidth]{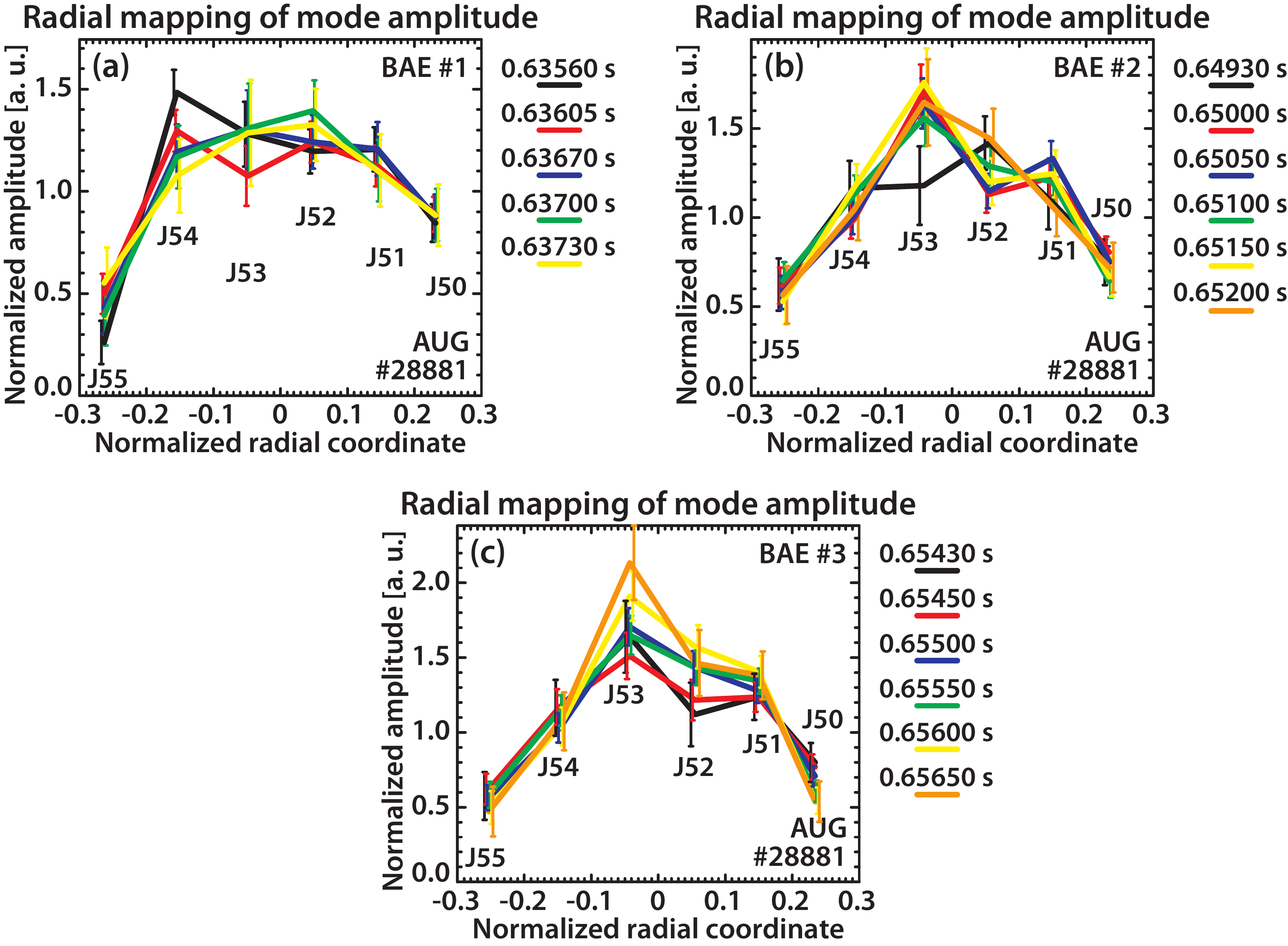}
  \caption{The radial mapping of the oscillation amplitude of BAEs. The results do not show significant changes in the radial distribution of the mode amplitudes.}
  \label{fig:bae_radmap}
\end{figure}

The time evolution of the instantaneous amplitude of chirp \#2 on the different SXR LOSs is shown in figure~\ref{fig:bae_amps_maxorder1}a.
In order the reduce the effect of the noise a smoothing is applied by a moving average with boxcar kernel of $0.5$ ms width.
The smoothed amplitudes are presented in figure~\ref{fig:bae_amps_maxorder1}b.
To examine the changes in the radial eigenfunctions, I constructed a radial mapping of the mode amplitudes.
This means that each LOS is labelled with the normalized poloidal flux of the magnetic flux surface to which the LOS is tangential.
The mode amplitude is calculated for each LOS and it is plotted as the function of the radial coordinate.
This way the radial mapping of the mode amplitudes can be evaluated at any time instant.
Since this work focuses on the relative changes in the radial eigenfunctions of the modes, each curve of the radial map was normalized with its integral.
This radial map for chirp \#2 is shown in figure~\ref{fig:bae_radmap}b at time instances indicated with pink arrows in figure~\ref{fig:bae_amps_maxorder1}.
Similarly, figure~\ref{fig:bae_radmap}a and figure~\ref{fig:bae_radmap}c show the time evolution of the mode amplitude distribution for chirps \#1 and \#3.
The error bars in figure~\ref{fig:bae_amps_maxorder1} and \ref{fig:bae_radmap} are derived from the estimated background noise level of the SXR spectrograms as it was described in section~\ref{sec:background}.
The results do not show significant changes in the radial distribution of the mode amplitudes.
This suggests that if there is any change in the radial eigenfunction it is smaller than the uncertainty of our measurement.
This observation is consistent with the physical picture that BAE is a normal mode of the plasma and its radial structure strongly depends on the background plasma parameters rather than on the EP distribution.

\section{Energetic particle driven GAMs}

EGAMs were observed in the same (or similar) discharges (see table~\ref{tab:shots}), but in different time-frequency intervals as BAEs.
I found 5 cases where the signal-to-noise ratio was appropriate for further analysis.
The strongest EGAMs were observed in discharge \#31213, where the angle of the NBI beam-line was most off-axis ($7.13^{\circ}$ respect to the horizontal axis).
In this discharge 3 consecutive EGAMs were found where the signal-to-noise ratio of the SXR signals was sufficient.
The evaluated fast ion distribution had a peak at $\rho_{\textrm{pol}} \approx 0.45$ according to the fast ion D-alpha (FIDA) spectroscopy measurements~\cite{geiger12fast, lauber14offaxis}.
$\rho_{\textrm{pol}} = \Psi/\Psi_0$ is the normalized poloidal flux which acts as a flux label in this case.

\subsection{Mode number analysis and localization of EGAMs}

\begin{figure}[htb!]\centering
  \includegraphics[width = 0.7\textwidth]{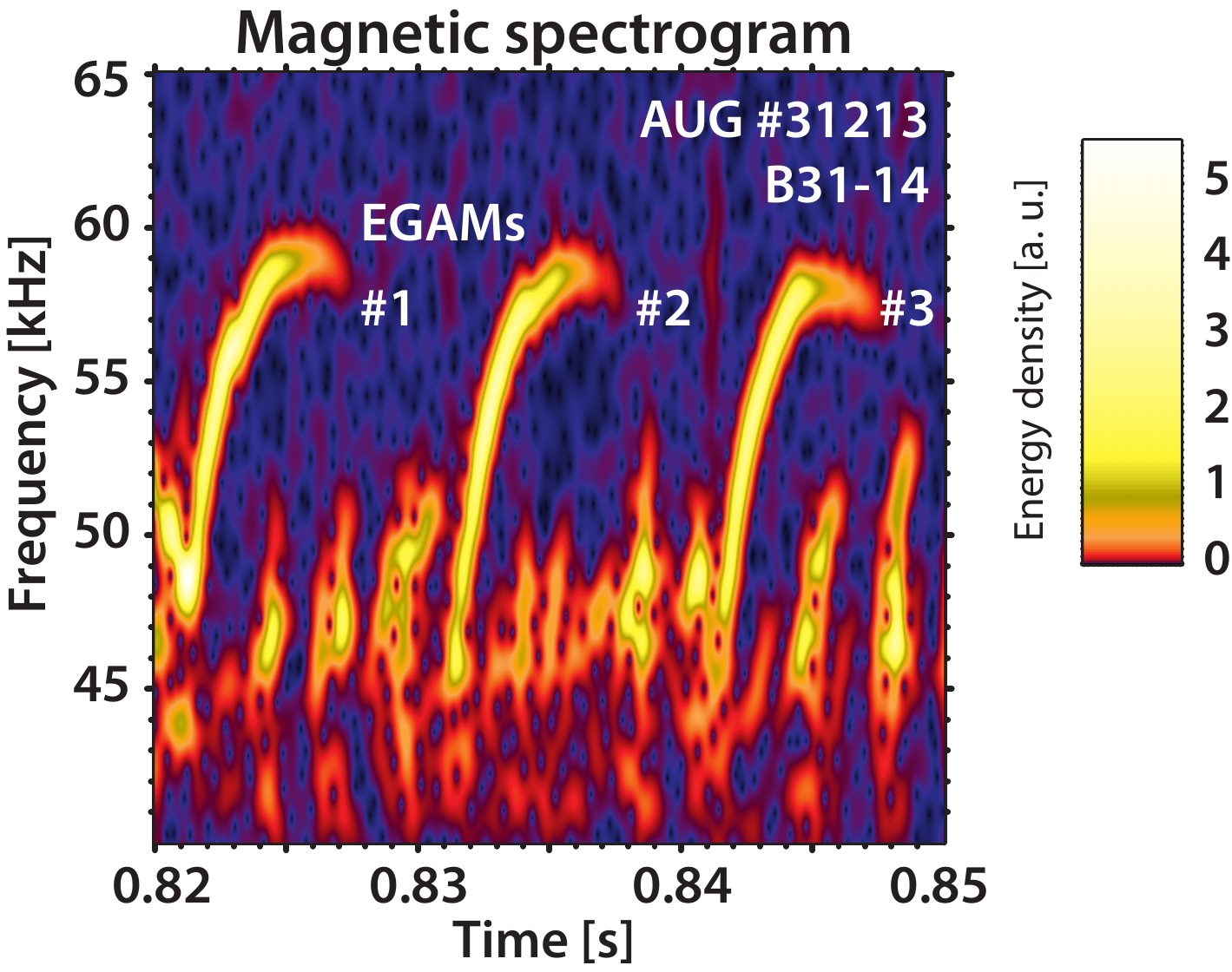}
  \caption{Chirping EGAMs with increasing frequency in the range of 45~-~60~kHz are visible on the magnetic spectrogram. Numbers denote the chirps which were investigated in detail.}
  \label{fig:egam_magnetic}
\end{figure}

For mode identification and mode number analysis again the magnetic pick-up coils were used.
The time evolution of the mode frequency was traced using the same ridge-following algorithm as for BAEs.
The magnetic spectrogram from discharge \#31213 which shows the investigated 3 consecutive EGAMs is presented in figure~\ref{fig:egam_magnetic}.

\begin{figure}[htb!]\centering
  \includegraphics[width = 1.0\textwidth]{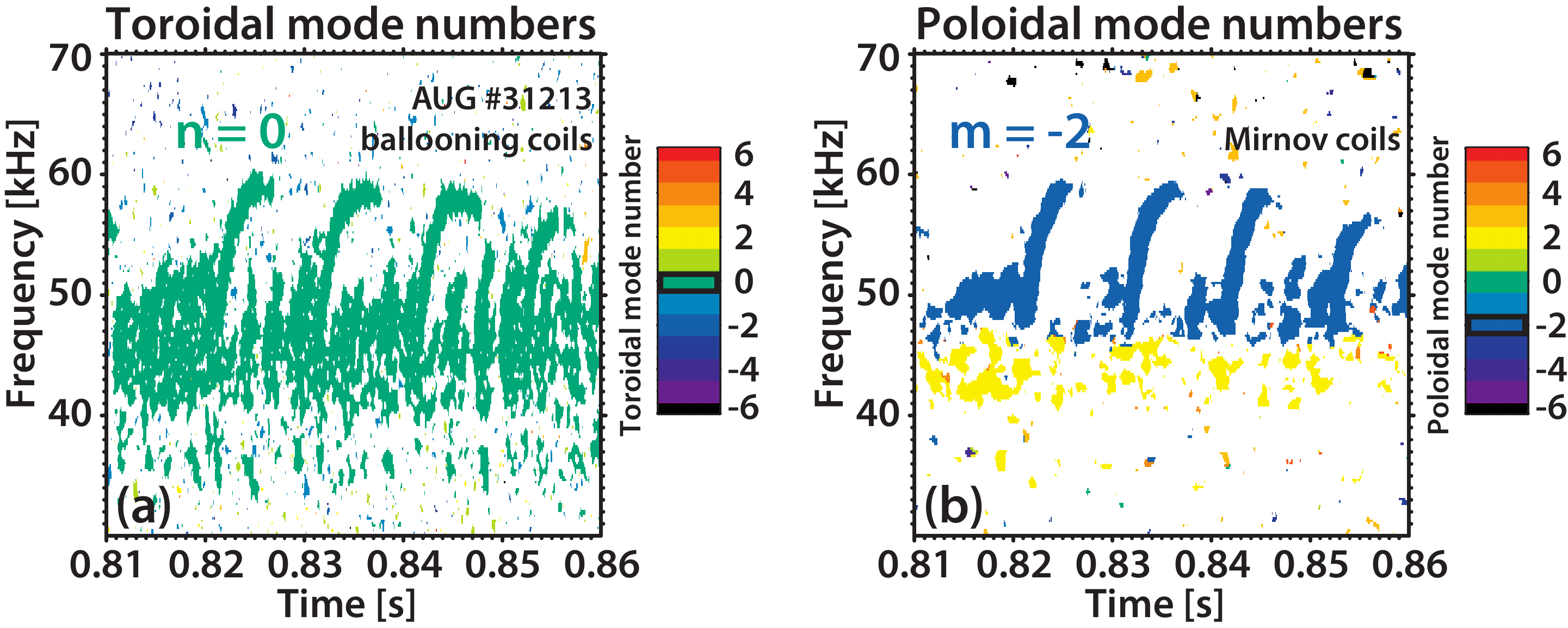}
  \caption{The result of time-frequency resolved mode number calculation. \textbf{(a)}~The toroidal mode numbers are plotted only in time-frequency points where the residual of the fit is lower than the 3~\% of the maximum. \textbf{(b)}~The poloidal mode numbers are plotted only in time-frequency points where the value of the minimum coherence is higher than~0.3.}
  \label{fig:egam_modenum}
\end{figure}

The mode number analysis of EGAMs was carried out in a similar way as it was done for the BAEs.
The results of the mode number evaluation are presented in figure~\ref{fig:egam_modenum}, where it is visible that all three EGAMs have $n=0$ toroidal and $m=-2$ poloidal mode number.

The next issue to be resolved is to determine the radial location of the mode.
One estimation of the radial location can be given from the SXR measurements.
Starting from the plasma core, the first SXR LOS on which the mode is not visible gives an outer boundary of the mode location.
Considering the SXR spectrograms the investigated EGAMs are located inside the $\rho_{\mathrm{pol}}=0.4$ surface.
The radial location of the mode was also estimated from the reflectrometry measurements.
Four channels measuring density fluctuations were used with fixed frequency (24, 33, 36 and 53 GHz) from the low field side.
The corresponding density values are $0.71$, $1.35$, $1.61$ and $3.48\cdot10^{19}$ $1/$m${}^{3}$.
The detected modes are visible on the spectrograms of channels with frequencies 33 and 36 GHz.
According to the density profile this means that the mode is located around $\rho_{\mathrm{pol}} \sim 0.25-0.45$.
Both results are consistent with the FIDA measurement from which it is expected that the strongest drive occurs closer to the core than where the fast ion distribution peaks ($\rho_{\mathrm{peak}} \sim 0.45$).

\subsection{Changes in the radial structure of EGAMs}

\begin{figure}[htb!]\centering
  \includegraphics[width = 1.0\textwidth]{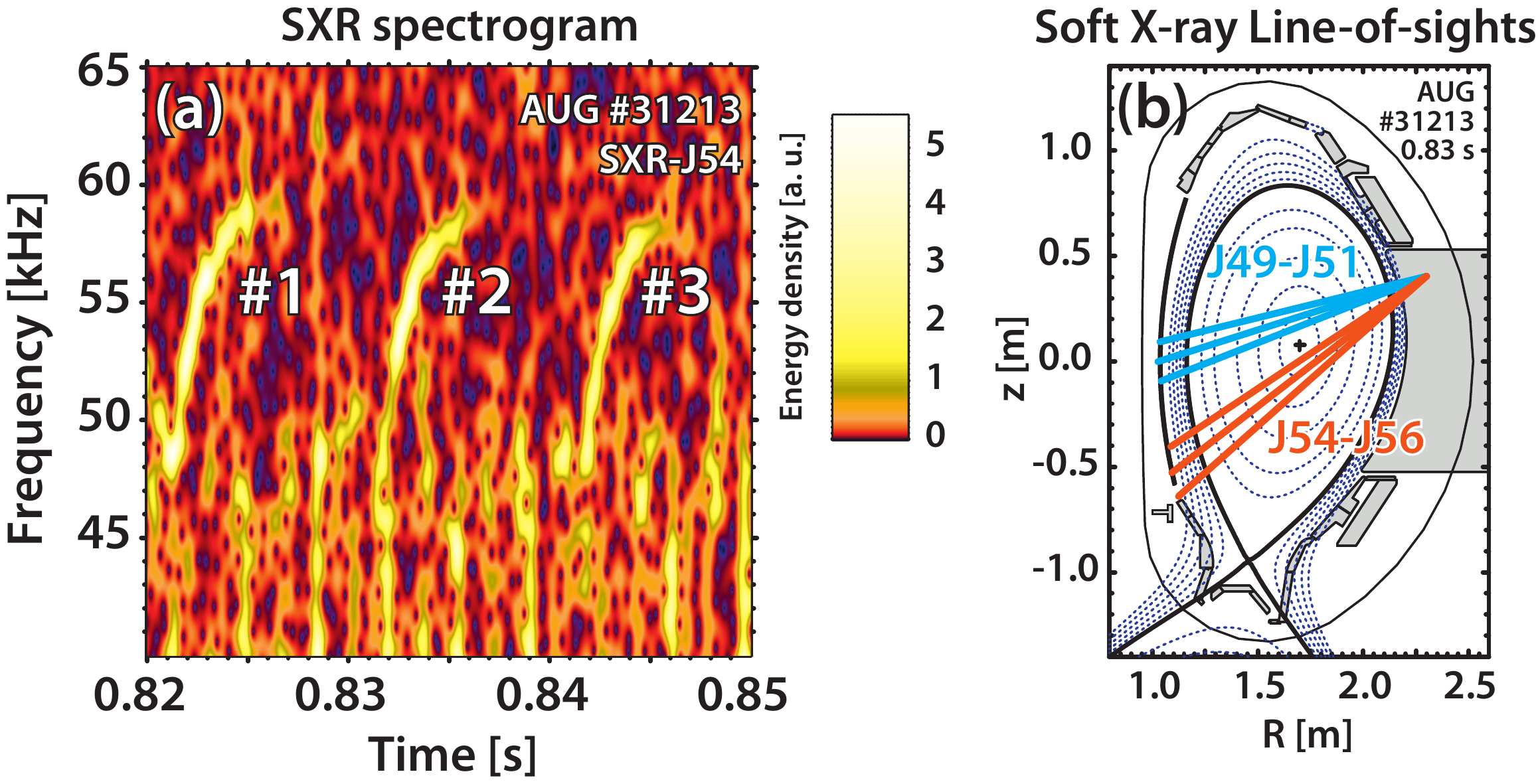}
  \caption{\textbf{(a)}~Soft X-ray spectrogram with 3 consecutive EGAMs. \textbf{(b)}~The position of the selected 6 SXR LOSs in a poloidal cross-section of AUG. These channels were used for the radial structure analysis of EGAMs.}
  \label{fig:egam_sxr_spectrogram}
\end{figure}

The time evolution of the radial structure of EGAMs was investigated in the same way as it was done for BAEs.
Again, the LOSs of SXR camera J were chosen.
The selected 6 LOSs where the signal-to-noise ratio was appropriate are shown in figure~\ref{fig:egam_sxr_spectrogram}b in a poloidal cross-section of AUG.
The soft X-ray spectrogram of channel J54 with 3 consecutive EGAMs is shown in figure~\ref{fig:egam_sxr_spectrogram}a.

\begin{figure}[htb!]\centering
  \includegraphics[width = 1.0\textwidth]{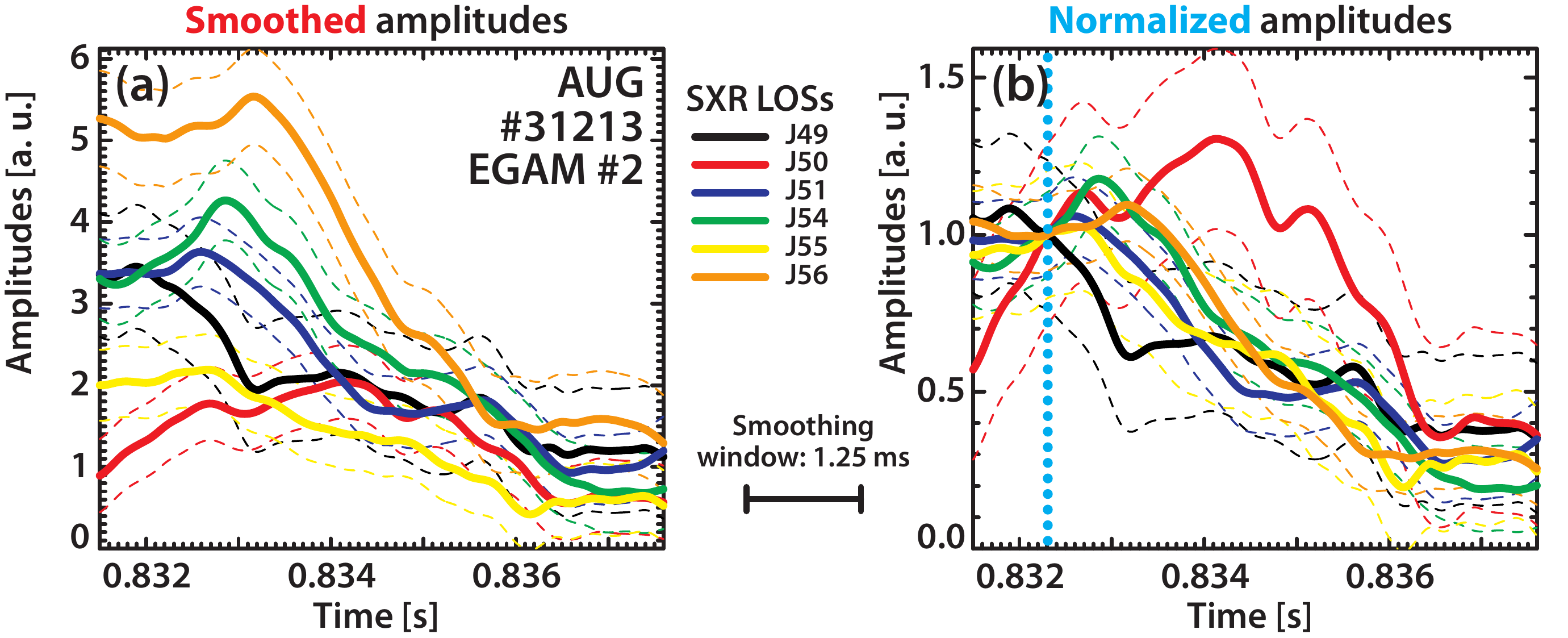}
  \caption{\textbf{(a)}~The time evolution of the oscillation amplitude of chirping EGAM \#2 (see figure~\ref{fig:egam_sxr_spectrogram}a) on the different SXR LOSs. The amplitudes were smoothed with a moving boxcar kernel of $1.25$ ms width. \textbf{(b)}~The amplitudes shown in figure~\ref{fig:egam_amps_maxorder1}a are normalized to the amplitude at the beginning of the chirp (indicated with dashed cyan line) for further analysis.}
  \label{fig:egam_amps_maxorder1}
\end{figure}

The amplitude of the mode is evaluated for all LOSs by using the formula defined in eq.~\eqref{eq:stft_linear_4}.
The reconstructed amplitudes of chirp \#2 (see figure~\ref{fig:egam_sxr_spectrogram}a) as a function of time are presented in figure~\ref{fig:egam_amps_maxorder1}a for channels shown in figure~\ref{fig:egam_sxr_spectrogram}b.
In this figure, the amplitudes were smoothed by a moving average with boxcar kernel of $1.25$ ms width.
The uncertainty of the result is indicated by the dashed lines.

It is already visible in figure~\ref{fig:egam_amps_maxorder1} that the time evolution of the amplitude is different on the different channels.
Since the SXR measurement is not well localized, the exact changes in the radial eigenfunction cannot be tracked, but its behaviour can be qualitatively described.
The amplitudes shown in figure~\ref{fig:egam_amps_maxorder1}a are normalized to the amplitude at the beginning of the chirp for further analysis.
The result is shown on figure~\ref{fig:egam_amps_maxorder1}b.
In order to investigate the changes of the radial structure, the radial mapping was evaluated, i.e. the normalized amplitudes are plotted as a function of the radial position of the channels in different time points.

\begin{figure}[htb!]\centering
  \includegraphics[width = 1.0\textwidth]{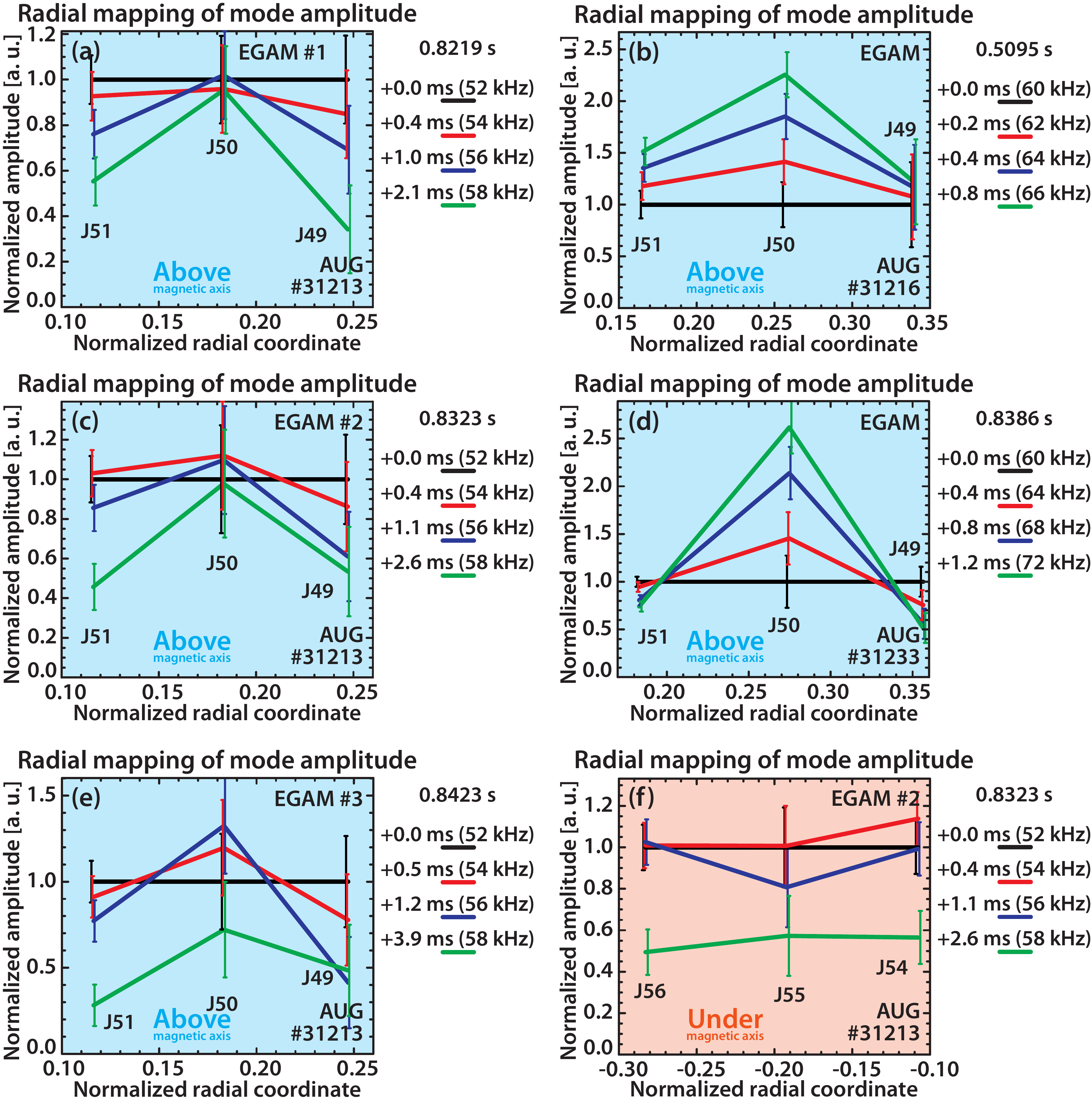}
  \caption{The radial mappings of the oscillation amplitude of EGAMs. \textbf{\mbox{(a-e)}}~5~cases showing the radial mapping from SXR LOSs located \emph{above} the magnetic axis. Shrinkage of the mode is visible in all cases. \textbf{(f)}~One case showing the radial mapping from SXR LOSs located \emph{under} the magnetic. Radial mappings calculated from channels under the magnetic axis do not show any significant change.}
  \label{fig:egam_radmap}
\end{figure}

The LOSs above and under the magnetic axis are handled separately and the time evolution of the radial mappings are shown in figure~\ref{fig:egam_radmap}.
On LOSs located above the magnetic axis (figure~\ref{fig:egam_radmap}a-e), a shrinkage of the mode is visible, since as time evolves the relative amplitude in the middle channel is rising compared to the outer channels.
One example for the radial mapping calculated from channels under the magnetic axis is presented in figure~\ref{fig:egam_radmap}f.
Regarding all 5 cases, radial mappings calculated from channels under the magnetic axis do not show any significant change.
This is most probably due to the longer distance from the observation point to the mode of these LOSs as it is visible in figure~\ref{fig:egam_sxr_spectrogram}b.
The LOSs located under the magnetic axis observe a bigger volume, detecting more light from the background plasma and relatively less from the flux surfaces where the mode is.
However, the shrinkage of the mode is visible in radial mappings calculated from channels above the magnetic axis.
As it is visible in figure~\ref{fig:egam_radmap}a-e, this shrinkage was significant in all 5 investigated cases.

\clearpage
\subsection{Discussion}

In the following section the theoretical explanation of the experimental results are presented.
The simulation results shown in this part are provided by Philipp Lauber~\cite{lauber15personal}.
The observed shrinkage in the mode structure of EGAMs during the non-linear chirping phase is consistent with our present theoretical understanding.
The mode coupled to the EGAM which was measured by the magnetic pick-up coils has a poloidal mode number $m = -2$.
However, the EGAM itself has a poloidal mode number $m = 0$.
The resonance condition to drive EGAMs simply follows from eq.~\eqref{eq:resonance_condition} if $n = 0$, $m = 0$ and $l = -1$ are substituted:
\begin{equation}\label{eq:egam_resonance}
  \omega - \omega_t = 0 ,
\end{equation}
where $\omega$ is the mode frequency and $\omega_t$ is the transit frequency of the interacting particles.
This EGAM resonance condition shows that the mode frequency is equal to the transit frequency of the interacting ions.

The exact phase space and real space coordinates of the wave-particle interaction are determined by several factors.
First, the EGAM drive is proportional to the velocity phase space gradient in the EP distribution function.
Second, the strength of the drive is proportional to the EP density.
In addition, the spatial dependence of the damping is an important factor.
Since in case of EGAMs, the main damping mechanism is ion Landau damping~\cite{nguyen09excitation}, the damping decreases with temperature.
Thus, the damping increases towards the plasma center while the EP density -- in the discharge \#31213 -- had an off-axis peak.
As a result, the radial position of the mode is expected to be closer to the magnetic axis than the peak in the EP density.
From the SXR measurements, the radial position of the mode is expected to be at $\rho\approx0.25$, thus the EP distribution was investigated at this position.

\begin{figure}[htb!]\centering
  \includegraphics[width = 1.0\textwidth]{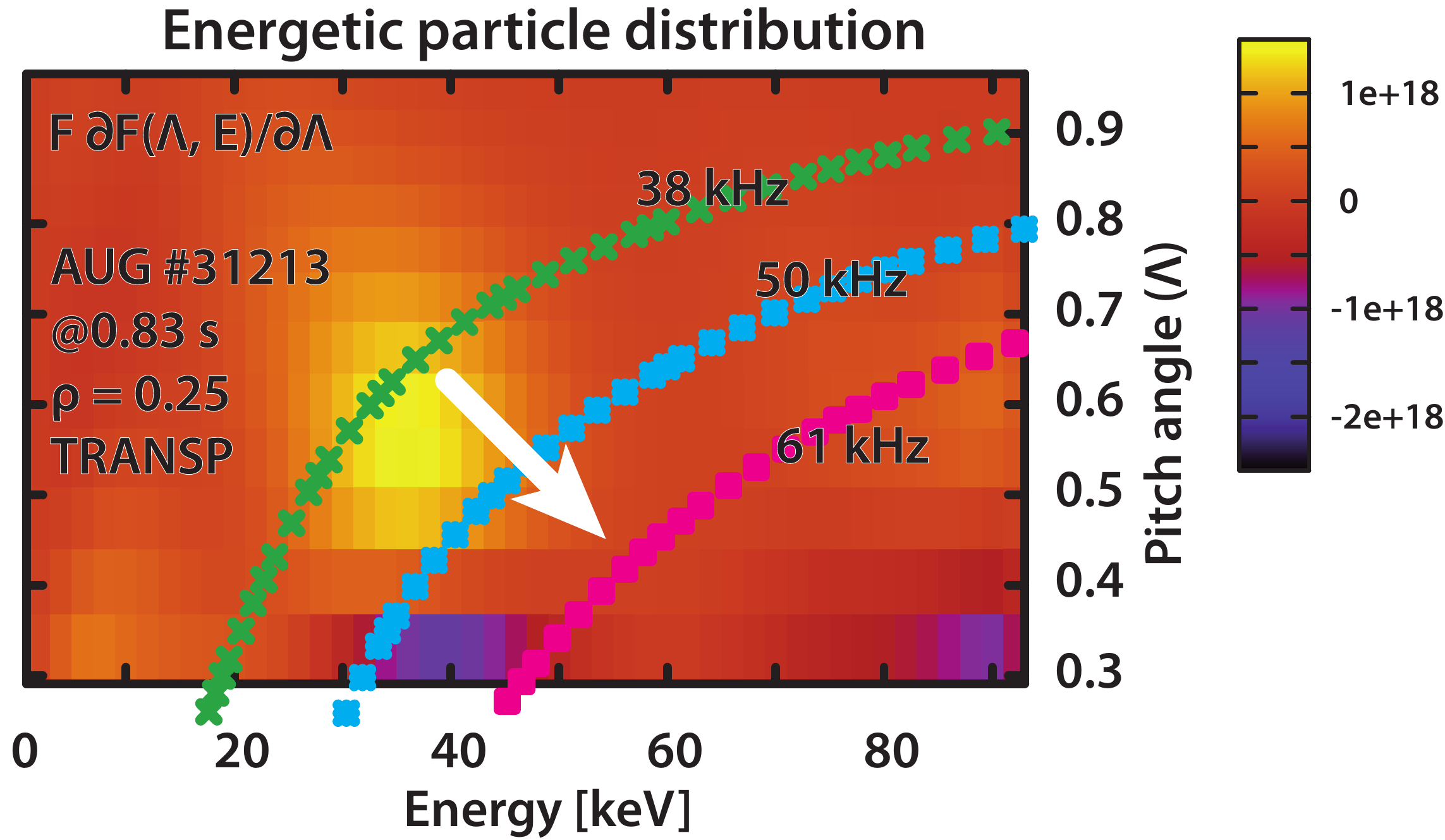}
  \caption{The contour plot shows the EP density ($F(\Lambda, E)$) times its derivative ($\partial F/\partial\Lambda$) as a function of the pitch angle ($\Lambda$) and energy ($E$) at $\rho\approx0.25$ in discharge \#31213 at $t=0.83$ s, where EGAMs were observed. The coloured curves show the $(\Lambda, E)$ coordinates of particles with different transit frequencies (38, 50 and 61 kHz). \textit{(This figure was created from data provided by Philipp Lauber~\cite{lauber15personal}.)}}
  \label{fig:egam_resonance}
\end{figure}

The EP distribution at $\rho\approx0.25$ was calculated by the TRANSP simulation code~\cite{lauber15personal, pankin04tokamak} at the time instance when EGAMs appeared.
The EP density ($F(\Lambda, E)$) times its derivative ($\partial F/\partial\Lambda$) is plotted in figure~\ref{fig:egam_resonance} as a function of the pitch angle ($\Lambda$) and energy ($E$).
The EP distribution can excite the mode when the drive overcomes the damping.
This can be reached where the drive is highest, i.e. at the peak of the $F\cdot\partial F/\partial\Lambda$ function which is around $(\Lambda = 0.75, E = 40$ keV$)$ this case as it is shown in figure~\ref{fig:egam_resonance}.
The coordinates of the peak of the $F\cdot\partial F/\partial\Lambda$ function determine the parameters of interacting particles which are passing particles.
The coloured curves in figure~\ref{fig:egam_resonance} show the $(\Lambda, E)$ coordinates of particles with different transit frequencies (38, 50 and 61 kHz).
The curves corresponding to higher frequencies move from the green curve ($38$ kHz) towards the pink curve ($61$ kHz) as it is shown by the white arrow.

\begin{figure}[htb!]\centering
  \includegraphics[width = 0.7\textwidth]{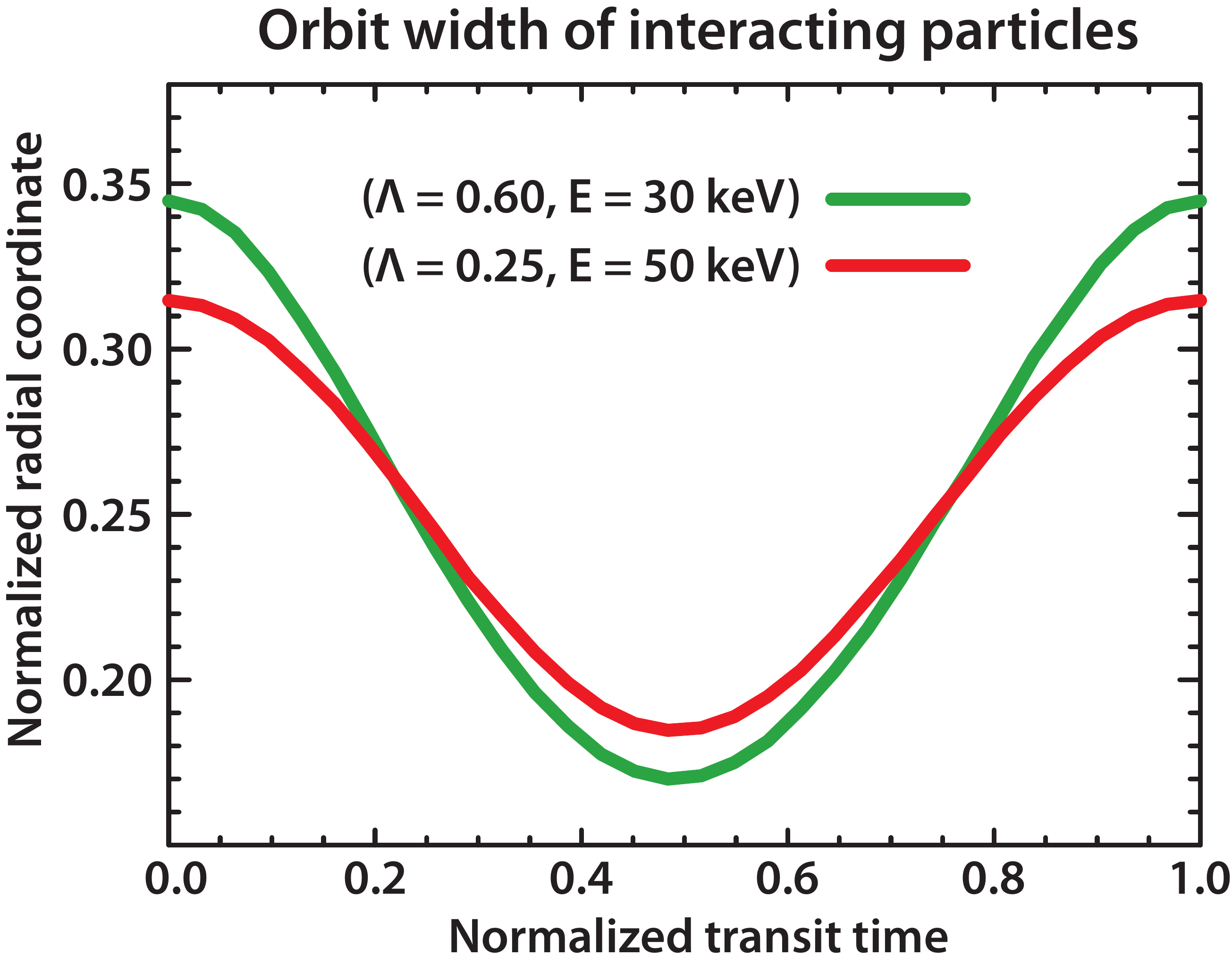}
  \caption{The orbit of EPs with different velocity phase space coordinates $(\Lambda, E)$. The radial coordinate of the particle is plotted as the function of the normalized circulation time. \textit{(This figure was created from data provided by Philipp Lauber~\cite{lauber15personal}.)}}
  \label{fig:egam_orbit}
\end{figure}

The mode frequency of the observed EGAMs starts at around $45$ kHz and increases until $60$ kHz as it is visible in figure~\ref{fig:egam_magnetic}.
The curve corresponding to the initial $45$ kHz frequency would intersect the region of the maximum peak in figure~\ref{fig:egam_resonance}.
This means that - considering eq.~\eqref{eq:egam_resonance} - the observed mode frequency is consistent with the simulated EP distribution function and the velocity phase space coordinates of the interacting particles are approximately $(\Lambda = 0.75, E = 40$ keV$)$.
The radial coordinate $\rho\approx0.25$ and the velocity phase space coordinates $(\Lambda, E)$ determine the orbit of the interacting particles.
The orbit of EPs is illustrated in figure~\ref{fig:egam_orbit} where the radial coordinate of the particle is plotted as the function of the normalized circulation time.
The green curve corresponds to the particles having $(\rho = 0.25, \Lambda = 0.60, E = 30$ keV$)$ coordinates and it shows that these interacting particles have a $\Delta\rho\approx0.34-0.17=0.17$ orbit width.
The mode structure of EGAMs is determined by the parameters of the drive, i.e. the orbit width of the interacting particles corresponds to the mode extent.

The mode frequency increases until $60$~kHz as it is visible in figure~\ref{fig:egam_magnetic}.
This means that as time evolves, the resonance condition changes and shifts to particles with higher transit frequency.
Figure~\ref{fig:egam_resonance} shows only the equilibrium EP distribution, the evolution of the EP density due to the non-linear interaction with fast ion cannot be followed.
However, a qualitative description can be given.
The pink curve in figure~\ref{fig:egam_resonance} shows the phase space coordinates of particles which have $61$ kHz transit frequency.
During the time evolution of the mode the resonance condition moves from $(\Lambda = 0.75, E = 40$ keV$)$ toward the pink curve.
This means that the resonance condition moves to the region of passing particles (particles with smaller $\Lambda$).
The exact time evolution of the resonance condition cannot be determined from the equilibrium EP distribution, but most probably the resonance follows the ridge in the $F\cdot\partial F/\partial\Lambda$ function indicated with the white arrow in figure~\ref{fig:egam_resonance}.
More passing particles have a narrower orbit width which can explain the experimentally observed shrinkage of the radial mode structure.
The orbit of EPs with coordinates $(\rho = 0.25, \Lambda = 0.25, E = 50$ keV$)$ is shown in figure~\ref{fig:egam_orbit} with red.
It is visible in this figure that the orbit width corresponding to the end of the EGAM time evolution ($\Delta\rho\approx0.31-0.18=0.13$) is narrower with approximately 25~\% than the initial orbit width at the beginning of the interaction.
The consequence of the narrower orbit width is the shrinkage of the mode structure, which is consistent with the experimental observations.

\chapter{Conclusions}

The understanding of energetic particle (EP) driven plasma modes plays a key role regarding future burning plasma experiments.
Super-thermal EPs in tokamak plasmas can excite various instabilities, and the most important transport process of EPs in the plasma core is their interaction with these global plasma modes.
The non-linear behaviour of the mode amplitude and frequency may exhibit a wide range of different behaviours, which significantly influences the impact of the instabilities on the fast particle transport.
Therefore, in order to comprehensively understand the non-linear behaviour of EP-driven instabilities, the investigation of these modes is essential.
In this thesis the rapid changes in the radial structure of beta induced Alfvén eigenmodes (BAEs) and EP-driven geodesic acoustic modes (EGAMs) were experimentally investigated during the non-linear chirping phase.

For the extensive characterization of the modes three diagnostic systems were chosen, namely the magnetic pick-up coils, the soft X-ray (SXR) cameras and the reflectrometry measurements.
In order to deal with the transient behaviour of the phenomena, short time Fourier transform was chosen as a basis of the data processing tools developed in this thesis.
The time evolution of the radial structure of EP-driven modes was examined by using the SXR measurements.
The SXR line-of-sights (LOSs) observe different parts of the plasma, which made it possible to analyse the spatial structure of the modes.

To follow the amplitude of the oscillation caused by the mode on the different LOSs, I developed an amplitude reconstruction method which can handle the rapidly changing mode frequency.
The derivation of the method and its validation with synthetic signals was presented in detail.
The effect of additive white noise on the reconstruction was also discussed in the thesis.
The uncertainty of the measurements was estimated from the properties of the background noise.

I investigated BAEs and EGAMs which were observed in the ramp-up phase of off-axis NBI heated plasmas in ASDEX Upgrade.
In order to identify the instabilities, a thorough mode number analysis was carried out.
Then the time evolution of their radial structure was examined.
The radial structure analysis showed that in case of the observed downward chirping BAEs the changes in the radial eigenfunction were smaller than the uncertainty of the measurement.
This behaviour is consistent with that the radial structure of BAEs -- as normal modes -- strongly depends on the background plasma parameters rather than on the EP distribution.

In case of rapidly upward chirping EGAMs the analysis consistently shows shrinkage of the mode structure.
This can be explained by the changing resonance condition in the velocity phase space of EPs.
The mode structure of EGAMs is sensitive to the EP distribution.
The rising frequency of the mode indicates that, as time evolves, the EGAM is driven by more passing particles which have narrower orbit width.
This leads to the experimentally observed shrinkage of the mode structure.

The results shown in the thesis will be presented at the 42nd EPS Conference on Plasma Physics~\cite{horvath15fast}.
My proposal to carry out further experiments in order to increase the event statistics of my investigation has been accepted.
Thus, dedicated discharges will be performed for further analysis of the examined chirping EP-driven modes in the subsequent campaign of the ASDEX Upgrade tokamak.

\chapter*{Acknowledgements}

I would like to express my special thanks to my supervisors Gergely Papp and Gergő Pokol for their continuous support of my research, for their motivation and enthusiasm.
I could not have imagined having better supervisors.
My sincere thanks also goes to Philipp Lauber.
This thesis would not have been possible to write without his help.

I would like to thank my collegues from BME-NTI, especially Gábor Pór and Péter Pölöskei for the fruitful discussions and the collegues from IPP Garching, especially Valentin Igochine and Marc Maraschek for their continuous help in diagnostic questions.

A special thanks to my family. Words cannot express how grateful I am to my mother, and father as well as my siter and my brother for all of the sacrifices that they have made on my behalf.
Finally I owe much to Eszti, without whose love I would not have completed this work.
\newline

This work has been carried out with support of FuseNet\footnote{\url{http://www.fusenet.eu}} -- the European Fusion Education Network -- within the framework of the EUROfusion Consortium.
This work has been carried out within the framework of the EUROfusion Consortium and has received funding from the European Union’s Horizon 2020 research and innovation programme under grant agreement number 633053. The views and opinions expressed herein do not necessarily reflect those of the European Commission.
I acknowledge the support of the Foundation for Nuclear Engineers (JNEA) and Hungarian State grant NTP-TDK-14-0022.

\clearpage

\clearpage
\phantomsection
\addcontentsline{toc}{chapter}{Bibliography}
\bibliographystyle{unsrt}
\bibliography{references}

\end{document}